# Design of an efficient Tunable Dual narrow-band MEMS Mid and Far IR emitter with Me-NTA for Industrial and Biomedical applications

Md. Imran Hasmi[1], Md. Saffat Gohor[1], Foez Ahmed[2], and Jaker Hossain[1*]

[1]*Photonics & Advanced Materials Laboratory, Department of Electrical and Electronic Engineering, University of Rajshahi, Rajshahi 6205, Bangladesh.*

[2]*Department of Information & Communication Engineering, University of Rajshahi, Rajshahi 6205, Bangladesh.*

**Abstract**

Spectrally selective infrared (IR) thermal emitters are gaining much attention now-a-days for sensing, spectroscopy and biomedical applications. In this research, two metasurface incorporated IR emitters are proposed and numerically analyzed using finite element method (FEM). First structure comprises a NiCr heater integrated with a NiCr-based metallic nanotube array (Me-NTA) metasurface to produce a single-narrowband emission in the mid-infrared (MIR) region. Furthermore, an Au-based Me-NTA metasurface on a NiCr-Au hybrid heater subsequently produces dual-narrowband emission in the short-and far-infrared (SIR and FIR) spectrums. Function of these emitters can be explained by Joule heating with the help of DC bias and consequently uniform temperature distribution can be observed along the active region. Simulation analysis shows that NiCr-metasurface based emitter produces single narrow-band near perfect emission centered at 4.5 μm in MIR region at an operating temperature of 700 K with maximum in-band conversion efficiency (CE) of 32.3% and radiated power of 199 mW. On the other hand, Au-metasurface based emitter generates dual-narrowband emission peaking at 2.5 μm and 10 μm, correlating to SIR and FIR subsequently, achieving maximum emission of 93% and 85%, respectively. The in-band CE for this emitter attains 10.4% and 4.4% in the first and second bands, associated with radiated powers of 350 mW and 147 mW, accordingly. Furthermore, execution of the emitter at 500 K reveals FIR emission with reduced power consumption. These results substantiate the possibilities of the suggested emitters in various industrial and biomedical applications.

**Keywords:** IR thermal emitters, Metasurface, NiCr Heater, Emissivity, MIR, FIR.

## 1. Introduction

Mid-infrared (MIR) spectrum, which usually lies within 2.5 μm to 20 μm that is known as "molecular fingerprint" region, has caught researchers' attention recently because of their vast application in molecular sensing and biomedical analyzing. This spectrum is further sectionalized into the short-wave infrared (SWIR, 1.4–3 μm) then the mid-wave infrared (MWIR, 3–8 μm) and lastly the long-wave infrared (LWIR, 8–20 μm), respectively. Wavelength from 8 μm to 1000 μm is also commonly known as far infrared (FIR) region. These regions individually contribute in different medical, environmental, industrial and security applications. MIR spectrum facilitates sensors to sense minute concentration of different molecules without substantial interference by fundamental rotational and vibrational resonances [1–3]. Considering the importance of this spectrum, IR emitters that can radiate electromagnetic waves in this region are becoming essential for sensing and imaging applications for biomedical and industrial purposes. For industrial uses, sensitive and selective gas measurements can be done using traditional IR emitters by emitting MIR or LWIR waves [4]. Also, for non-invasive biomedical spectroscopy and analyzing, MIR spectra specially SWIR is very much applicable [3]. Additionally, FIR emitters that emit "growth rays" ranging from 4 μm to 14 μm can be used for therapeutic applications like recovering damaged cells and strengthening blood circulation [5]. As a consequence, different types of IR emitters have been proposed by the researchers throughout the years. Conventional IR emitters produce broadband EM wave across mid to far infrared wavelengths by using thermal blackbody-type sources, for example SiC globars or heated filaments. These broadband emitters have their particular applications because of their simplicity and stability but they display absence of spectral selectivity which is necessary for distinct sensing applications. Furthermore, constrained energy density per wavelength trait of broadband radiation and incompatibility for the applications that require highly directional, narrowband MIR or FIR emission obliged researchers to find supplementary approaches to produce narrowband MIR and FIR radiation [2].

An object can emit electromagnetic energy along its surroundings if the temperature of that object rise above absolute zero. This phenomenon is known as thermal radiation and it has been extensively studied by researchers for utilizing it in wide range of applications. Amongst these applications, thermal emitter plays a crucial role for gas-sensing systems which detects hazardous, explosive or toxic gases in both industrial and civil environments [6,7]. For addressing such applications, microelectromechanical system (MEMS)-based thermal emitters have shown promising prospects for their compact size, low thermal mass and rapid thermal response [8,9]. MEMS emitters low thermal mass which is driven by short time-based Joule heating of the microheater leads to a thermal radiation. However, this thermal radiation could be of broadband in nature if it's a conventional thermal emitter which employs traditional blackbody thermal emission in the IR wavelength range [10]. This broadband nature is undesirable in some cases such as selective wavelength emitter [11]. Therefore, spectral engineering became eminent to produce selective narrowband emission.

It has become challenging to manipulate thermal radiation by natural materials or approaches and has been a substantial research topic for many years. According to the Kirchhoff's law, in thermal equilibrium condition, the absorptivity of an arbitrary body is equivalent to the emissivity which is the core concept for controlling thermal radiation. Over the last few years, advancement in research facilitated manipulation of thermal radiation by introducing artificial micro or nano structures and they have become a pioneer for tailoring light absorption. These artificially engineered structures, better known as metamaterials, are an outstanding applicant to overcome those before-mentioned challenges and selective wavelength emission [12–14]. To manipulate electromagnetic waves, material comprising distinctive optical property is a must which cannot be found in natural occurring materials. To realize this, researchers expanded the boundary of possibilities and engineered such subwavelength micro or nano structures arranged in periodic pattern that exhibit supernormal properties such as negative refractive index, reverse Doppler effect, perfect absorption etc. By fine tuning of geometrical dimensions and material compositions of their subwavelength periodic patterns, the permittivity and permeability can be tailored to achieve these unconventional attributes. Many applications utilize these studies of metamaterial for sensitive gas sensing, energy harvesting, medical imaging along with thermal emitters [14–30]. Furthermore, tunability of geometrical dimensions of metamaterial plays a vital role in order to regulate operating wavelength region. For instance, visible [31,32], IR [33,34], terahertz [35–37], microwave light [38]. Beyond geometrical tailoring, integration of metamaterials with MEMS technology can deliver desirable setting for achieving selective wavelength emission. Additionally, implementation of microheater and metamaterials can be impactful for high efficiency and tunable radiation intensity by the altering quantity of heating energy generated from microheater [39]. Moreover, for spectral engineering of thermal radiation, metallic nanotube array (Me-NTA) type structures are introduced in these emitters. This structure substantiates localized surface plasmon and cavity-type resonance that helps to emit confine and wavelength-selective infrared wave. Therefore, broadband radiation gets suppressed and preferable emission peak for sensing applications. Recent studies have showed potential prospectives for Me-NTA structures for narrowband and tunable thermal emission in mid and far infrared region [40,41]. Numerous studies demonstrated dual or multi band thermal emission incorporating plasmonic and metamaterial absorbers however most of them focused on gas sensing and thermal regulation applications. On the other hand, most reported MEMS-based thermal emitters demonstrated broadband emission [9,42–44]. Thus, a single framework for achieving efficient and tunable narrowband emission comprising both metamaterial and MEMS technology and utilizing Me-NTA structure has become obligatory for application specific narrowband infrared emission which is a novel approach in this area. A common yet profound material Nichrome (NiCr) is used as both heater and Me-NTA structure in one of the proposed emitters due to its high resistivity and thermal stability with minimal power consumption that is also compatible for MEMS fabrication [44]. Although NiCr is not a classical plasmonic metal but it can facilitate infrared resonance by integrated as Me-NTA structure. Furthermore, a thin layer of Au added onto NiCr can serve as the heater which reportedly increased the emission by enabling enhanced electrical contact. However, most NiCr based thermal emitters generate wideband blackbody emission [45]. Thus, NiCr-based narrowband thermal emitter has become a challenging yet attractive area of research to ensure effective spectral utilization. The Me-NTA structure comprises Au has strong surface plasmon resonance [11]. A widely used dielectric spacer known as Silicon carbide (SiC) displays unique combination of optical and thermal properties and can be used between the heater and Me-NTA structure. It enables surface phonon polariton (SPhP) modes in MIR that enhances emission when incorporated with emitter [46]. It can also provide robustness and electrical insulation for the emitter that is vital for high thermal applications.

Narrowband thermal emitters are necessary for specific applications such as gas sensing and biomedical therapy because they require wavelength-selective infrared (IR) radiation. Gas molecules absorb IR radiation at characteristic wavelengths respective to their vibrational modes in nondispersive infrared (NDIR) gas sensors. Significant energy loss take place in conventional NDIR systems because they utilize broadband thermal source where most of the emitted radiation remains unused. Thus, metamaterial based thermal emitters have been developed to improve efficiency by incorporating wavelength-selective emission. The 2.5 μm wavelength region coincides with absorption bands of $H_2O$ and various hydrocarbon gases. Additionally, 4.26 – 4.5 μm corresponds to the absorption band of $CO_2$, which make them essential for NDIR gas sensors [47,48]. Emission at 10 μm wavelength range can be employed for thermotherapy and tissue heating application [49].

Herein, two MEMS models are suggested as heater where one comprises only NiCr and other comprises an ultra-thin Au layer onto the NiCr surface. SiC is used as dielectric spacer for both structures. Lastly, NiCr Me-NTA structure is used as resonating layer for one and Au Me-NTA structure is used for another models. First one demonstrates a single narrowband with more than 90% emission. Second model displays dual narrowband emission peaks at 2.5 μm and 10 μm which can be used as MIR and FIR emitters. Top resonator layer is tailored such a way that it becomes independent of polarization as the electromagnetic wave generated from the heater have multiple polarization directions. For this reason, a uniform radiation power generation becomes possible. By careful optimization of various parameter of the propose model, a tunable single and dual narrow-band thermal emitter has been achieved.

## 2. Design and methodology

Two MEMS emitter configurations one is at MWIR and other cover both SWIR and FWIR are proposed. One configuration integrates single-layer heater with metasurface-based resonator while other incorporates double layer heater. Fig. 1 illustrate both devices configurations. The single-layer heater consists of a rectangular-shaped NiCr-based heater which is deposited on a silicon substrate to provide overall support for the device. A DC bias voltage is driving the heater which according to the conventional Joule effect, generate heat energy. Two silver electrodes have been placed at the two ends of the rectangular hot plate so that electrical energy can flow and as a result get converted into thermal energy. At this stage without tailored metastructure, electromagnetic spectrum produced by the thermal radiation power becomes broadband in nature. The heater is covered with a SiC dielectric layer which offers electrical insulation, thermal distribution and minimal conduction losses. A NiCr-based Me-NTA metasurface structure is fabricated on top of the dielectric layer to enhance narrowband thermal emission. However, a slightly different approach is taken for the double-layer heater model. Here, an additional Au-based hotplate is placed on top of the NiCr heater layer. This framework has been done in order to get a better quality of optical response including improved plasmonic confinement and better spectral selectivity from the device in FIR region. Other than that, this proposed structure also follows a SiC dielectric layer. Another modification compared to the single-layer heater device is that, this structure comprises an Au-based Me-NTA shaped metasurface.

For the NiCr-metasurface based IR emitter, the silicon substrate on which the device is constructed has a 0.2 mm thickness with a dimension of 12 mm x 5 mm. To assure electrical connectivity, silver (Ag) contact pads are considered having 0.45 µm thickness, 0.6 mm long and 5 mm wide. The single-layer NiCr heater is 0.3 µm thick having 10.8 mm x 5 mm dimension, resulting in an emitting area of 54 $mm^2$. Following that, the SiC dielectric layer comprises the thickness of 0.2 µm with same dimension as NiCr heater. Furthermore, for Au-metasurface based IR emitter, the thickness of the Si substrate is 0.5 mm, length is 30 mm and width is 20 mm. Ag contact pads have 0.45 µm thickness, 2 mm length and 20 mm width. Here, NiCr heater is 0.3 µm thick with a length of 26 mm and width of 20 mm. It also incorporates a thin 0.01 µm Au layer as secondary heater layer with dimensions same as NiCr heater. Thereafter, SiC dielectric layer consists 0.16 µm thickness again with equal dimension as NiCr heater. Thus, a 520 $mm^2$ of operational area is constructed for emission. Table 1 summarizes the structural details for both single- and double-layer heaters. The NiCr-based metasurface of single band IR emitter comprises hollow ring resonators in a periodic array. It has a periodicity of 2.5 µm, disk outer radius of 1 µm, disk inner radius of 0.9 µm and disk height 0.1 µm. On the other hand, Au-based metasurface of double band IR emitter has a periodicity of 2 µm, disk outer radius of 0.57 µm, disk inner radius of 0.5 µm and disk height 0.1 µm. Strong plasmonic resonance at desired IR wavelengths is achievable via these geometrical parameters. Table 2 summarizes the metasurface unit cell parameters.

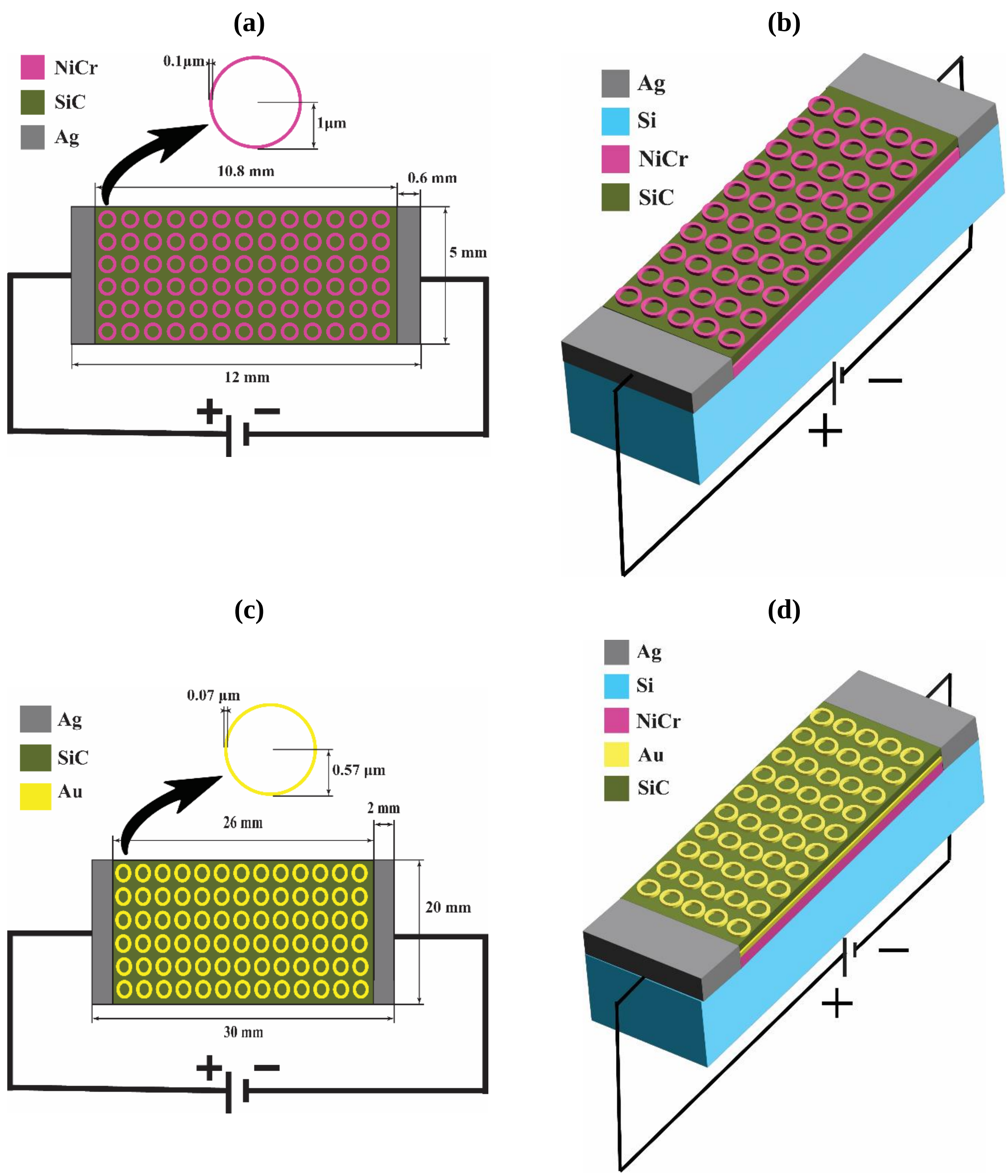

**Fig. 1:** Proposed NiCr Me-NTA metasurface-based IR emitter (a) top view, (b) 3D schematic view and Au Me-NTA metasurface-based IR emitter (c) top view, (d) 3D schematic view.

For microheater heating element, $Ni_{80}Cr_{20}$ was chosen as primary heating element due to its superior property such as high resistivity, low TCR and high tensile strength thus making it a worthy choice for electrically driven joule heat [50–52]. Additionally, its optical property does not vary much with thickness which makes it versatile for using as a thermal emitter for various thickness [12,53]. NiCr and Au material were modeled using Drude-Lorentz dispersion with corresponding plasma and collision frequencies. For NiCr, plasma frequency, ωp = $2.9\times10^{15}$ rad/s and collision frequency, $\gamma = 2.61\times10^{15}$ rad/s [54,55]. In order to improve overall selectivity and plasmonic response at desired wavelength, a thin gold layer is incorporated in dual band IR emitter having plasma frequency, ωp = $1.37\times10^{16}$ rad/s and collision frequency, $\gamma = 4.08\times10^{13}$ rad/s [56]. Other electrical, thermal and optical properties of materials such as NiCr, Au, Ag and SiC were collected from relevant literatures which are summarized in Table 3.

The electrothermal behavior of the proposed heaters was analyzed using COMSOL Multiphysics 6.3 software, by Joule heating module [57]. All metal-dielectric interfaces adopted continuity boundary conditions. Substrate edges acquired thermal insulation. Optical simulation was conducted with the Radio Frequency module (electromagnetic waves, frequency domain) which solved Maxwell's equations with periodic boundary conditions. Absorption was derived from $S_{11}$ and $S_{21}$ parameters which are reflection and transmission coefficient, respectively. Furthermore, Kirchhoff's law was used to correlate absorption (A) with emissivity (E) where, E = A [58].

In order to know the radiation characteristics of the proposed IR emitters, several steps have been taken into consideration including spectral emittance, band emissivity and integral net radiative heat flux for radiant power calculation. The radiant power which is commonly referred as opto-electro conversion from emitters top surface to another surface (vacuum as consideration), has been validated by a computational radiative heat transfer approach using COMSOL. The RHT in corporate with Surface-to-Surface radiation module, is the distribution of hemispherical emissive power $E_{b\lambda}$, described as the energy radiated by a black surface per unit time, per unit area, and per unit wavelength interval said Planck. The spectral emittance can be calculated by equating hemispherical emissive power for refractive index, n = 1 as follows [59]:

$$E_{b\lambda} = \frac{2\pi hc^2}{\lambda^5 \exp(hc/k\lambda T-1)} \quad \text{-----------------------------------------------(1)}$$

Where, h is the Planck's constant, c is the speed of light in vacuum, k is the Boltzmann's constant. The mean band emissivity for per band radiative heat flux, ε over wavelength $[\lambda_1, \lambda_2]$ at temperature T can be determined using spectral emissivity and the blackbody radiation as:

$$\varepsilon = \frac{\int_{\lambda 1}^{\lambda 2} \varepsilon_\lambda E_{b\lambda}(T) d\lambda}{\int_{\lambda 1}^{\lambda 2} E_{b\lambda}(T) d\lambda} \quad \text{----------------------------------------------------(2)}$$

where, $E_{b\lambda}$ is the blackbody spectral emittance and $\varepsilon_{\lambda}$ is the emitter spectral emissivity [58]. The Stefan-Boltzmann law shown in equation (3), interpret that the total emissive power of a blackbody is the collective radiation over all wavelength, emitted into a vacuum [59]:

$$E_b = \int_0^{\infty} E_{\lambda b}\, d\lambda = \sigma T^4 \text{-------------------------------------------(3)}$$

$E_b$ = blackbody hemispherical total emissive power, σ = Stefan-Boltzmann constant = 5.670 × $10^{-8}$ $W/m^2K^4$ and T = temperature. It is noteworthy that COMSOL automatically considers emitter surface and surrounding vacuum as interacting surfaces while considering surface to surface radiation. Moreover, the heat exchanged between two diffuse surfaces, I and J, with constant surface emissive powers referred to as net radiative heat exchanged is described as follows [59]:

$$Q_{IJ} = \frac{E_{Ib} - E_{Jb}}{\frac{1-\varepsilon_I}{\varepsilon_I A_I} + \frac{1}{A_I F_{IJ}} + \frac{1-\varepsilon_J}{\varepsilon_J A_J}} \text{---------------------------------------------(4)}$$

where, $Q_{IJ}$ = net heat exchanged, $E_{Ib}$ = hemispherical total emissive power of surface I, $\varepsilon_I$ = surface emissivity of I, $A_I$ = geometric area of surface I, $F_{IJ}$ = view factor from surface I to surface J.

**Table 1.** Structural parameters of single- and double-layer heater and substrate

| Device | Layer | Material | Thickness (μm) | Length (mm) | Width (mm) | Operational Area($mm^2$) |
|---|---|---|---|---|---|---|
| **Single band** | Substrate | Si | 200 | 12 | 5 | - |
| | Pads | Ag | 0.45 | 0.6 | 5 | - |
| | Heater | NiCr | 0.3 | 10.8 | 5 | 54 |
| | Dielectric | SiC | 0.2 | 10.8 | 5 | - |
| **Double band** | Substrate | Si | 500 | 30 | 20 | - |
| | Pads | Ag | 0.45 | 2 | 20 | - |
| | Heater 1 | NiCr | 0.3 | 26 | 20 | 520 |
| | Heater 2 | NiCr+Au | 0.3+0.01 | 26 | 20 | - |
| | Dielectric | SiC | 0.16 | 26 | 20 | - |

**Table 2.** Metasurface unit cell parameters.

| Parameter | Single-band | Double-band |
|---|---|---|
| Periodicity (μm) | 2.5 | 2 |
| Disk outer radius (μm) | 1 | 0.57 |
| Disk inner radius (μm) | 0.9 | 0.5 |
| Disk height (μm) | 0.1 | 0.1 |
| Material | NiCr | Au |
| Dielectric | SiC | SiC |

**Table 3.** Properties of materials incorporated in the devices.

| Material | σ (S/m) | k (W/m.k) | ρ (kg/m$^3$) | εr/model | References |
|---|---|---|---|---|---|
| NiCr | 9.09x10$^5$ | 17 | 8300-9000 | Drude/Lorentz | [60,61] |
| Au | 3.45x10$^7$ | 317 | 19,320 | Drude | [62,63] |
| Ag | 6.3x10$^7$ | 415 | 10,492 | COMSOL default | [64] |
| SiC | - | 5.8-7.8 | 3210 | εr = 10.8, tanδ = 0.003 | [65,66] |

## 3. Results and discussion

### 3.1 Electro-thermal characteristics of the MEMS heaters

#### 3.1.1 Temperature distribution of the heaters

Fig. 2 illustrates the temperature distribution at the surface of both proposed heaters. Firstly, in Fig. 2(a), NiCr heater is simulated at 2.23V and the resulting temperature distribution is achieved. From the figure the temperature distribution can be seen almost persistent at the center of the heater surface. This uniformity across the surface provides spectral stability which is crucial for consistent emission. Minimal gradient of the temperature can be seen at vertical edges of the surface. This anisotropic phenomenon can be elucidated by the fact that heat is conducted away more rapidly near the perimeter of a heater. Also, vertical edges of the rectangular heater displayed more efficient heat loss compared to horizontal edges as center of the surface retains more heat than edges [67]. However, similar temperature description suites for the NiCr-Au heater simulated at 2.67 V as depicted in Fig. 2(b).

**(a)** **(b)**

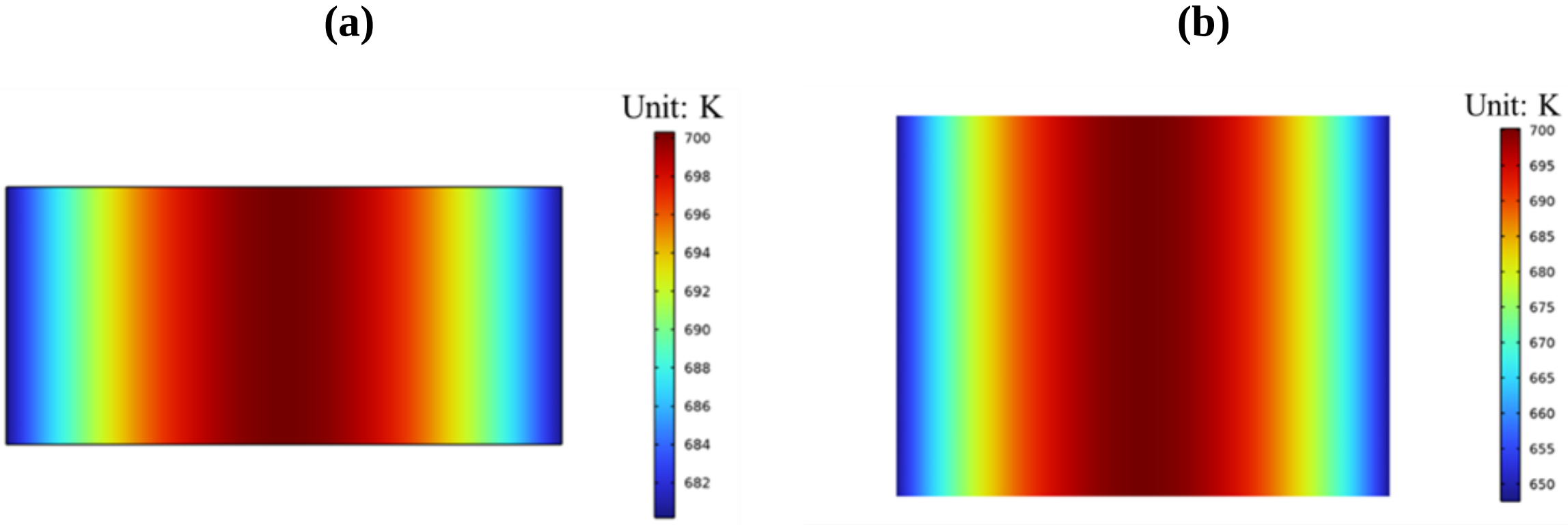

**Fig. 2:** Surface temperature distribution of (a) NiCr heater simulated at 2.23V and (b) NiCr-Au heater simulated at 2.67 V.

#### 3.1.2 Voltage-dependent temperature effect of the heaters

In Fig. 3, the temperature characteristics according to voltage has been shown for both proposed heaters. From the figures, a non-linear quadratic relationship is observable between voltage and temperature. Fig. 3(a) and (b), varying voltage and the corresponding temperature can be also seen to increase quadratically. Joule heat $Q_{in}$ can be expressed as [68]:

$Q_{in} = (V_{in}^2/R) \int t dt$ --------- (5)

Where, $V_{in}$ is applied voltage and R is the resistance. Therefore, for a given resistance, higher voltage generates higher temperature. The optimum voltage is selected 2.23V for NiCr heater for stable operation and it generated a temperature of 700K.

Heater comprised with NiCr and Au also exhibits similar temperature voltage relation. At optimum 2.67V NiCr with Au heater generated 700K.

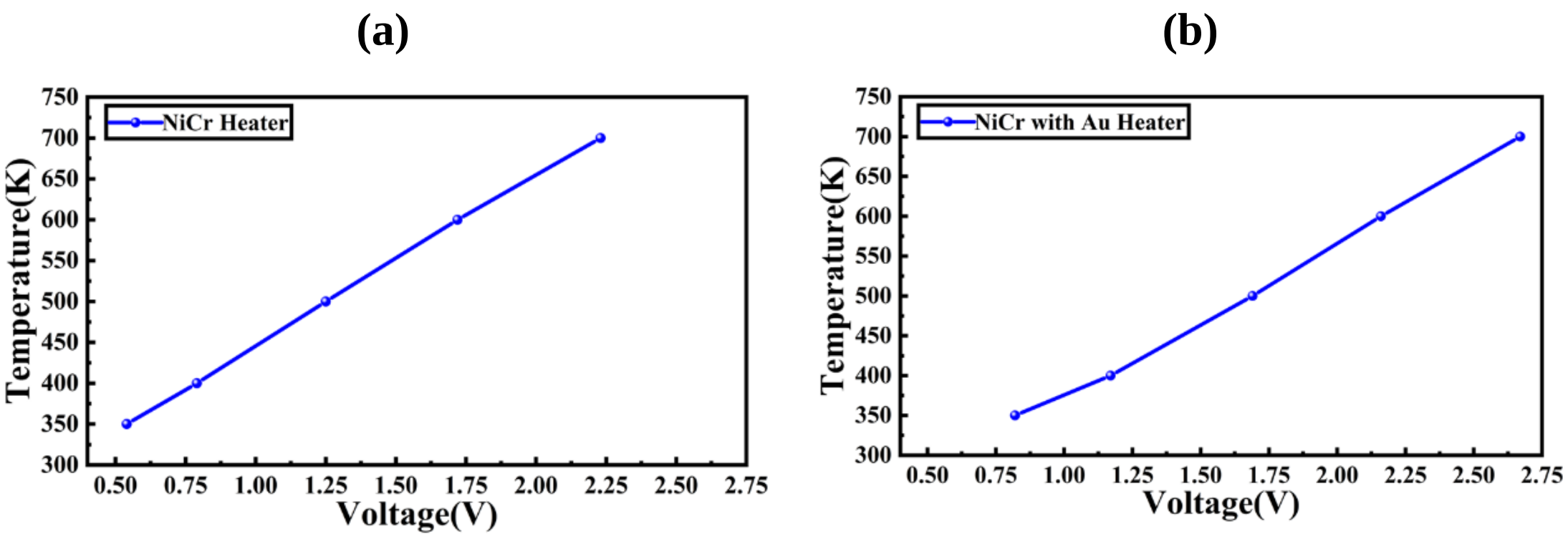

**Fig. 3:** Temperature variation with input voltage of (a) NiCr heater and (b) NiCr-Au heater.

### 3.1.3 Input power characteristics of the heaters

In the Fig. 4, consumed power is illustrated as a function of input voltage for both heaters. The power curve demonstrated a parabolic relationship with the input voltage. The power consumption increased with the increasing voltage in both cases. For NiCr heater consumes 615mW at optimum 2.23V as depicted in Fig. 4(a) it whereas for NiCr-Au heater it consumes 3.33W at 2.67V as delineated in Fig. 4(b).

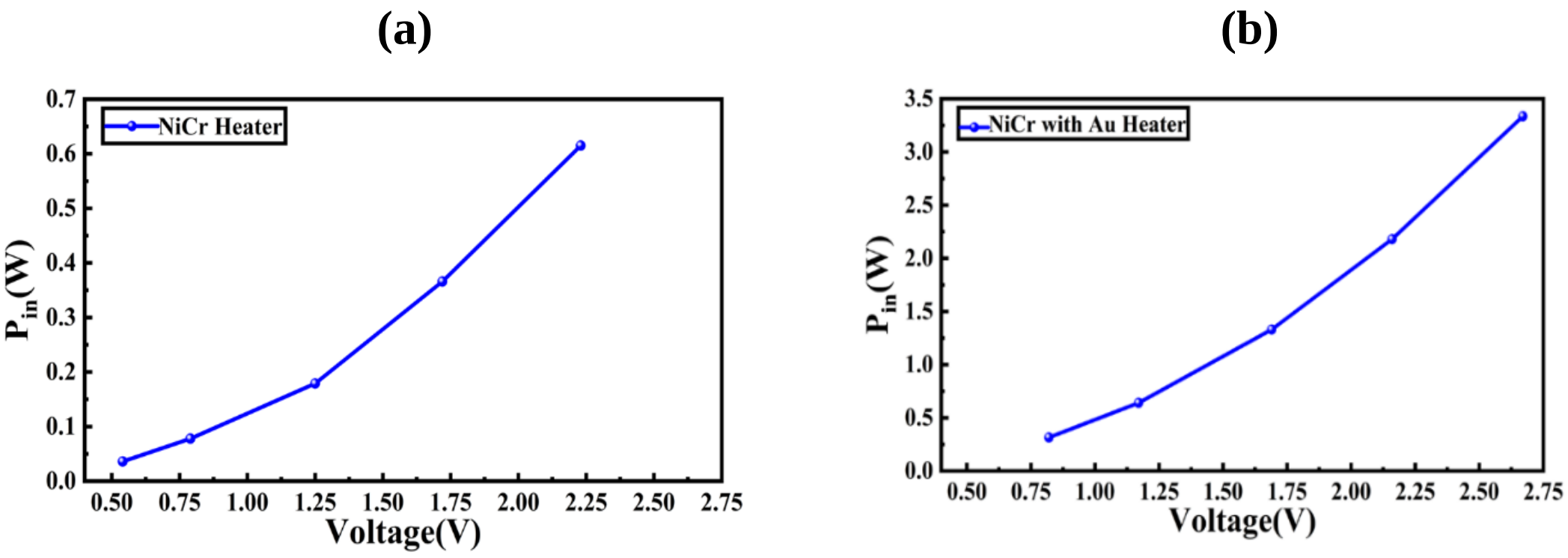

**Fig. 4:** Voltage dependent input power characteristic of (a) NiCr heater and (b) NiCr-Au heater.

### 3.2 Parametric and performance analysis of the NiCr Metasurface-based single band IR emitter

For achieving the distinctive narrowband emissions in the MIR and FIR band, a methodical parametric analysis of the suggested metasurface based IR emitters is carried out. The parametric study includes variation in outer disk radius and height of the Me-NTA surface, dielectric spacer thickness and ground plane thickness. These evaluations demonstrate how the changes in these above-mentioned parameters alter the emissivity spectrum within the operational spectrum. Also, different performance attributes such as emissivity, spectral emittance, voltage and power analysis, efficiency etc. are discussed in this section.

#### 3.2.1 Parametric analysis of NiCr Metasurface-based single band IR emitter

Fig. 5(a) depicts the impact of NiCr Me-NTA disk outer radius variation on the emissivity spectrum. Disk outer radius is varied from 0.8 µm to 1.2 µm across 1 µm to 15 µm wavelength and corresponding emissivity is recorded. It is clear from the figure that negligible change occurs within the spectrum and 98% emissivity is achieved at 4.5 µm. This negligible impact on emissivity spectrum of disk outer radius variation can be explained because of relatively high intrinsic ohmic loss of NiCr material in the MIR region which has been used as metasurface for the first proposed design. Therefore, for lossy materials like NiCr, emission is governed by material damping more and slight geometric changes impose minimal effect [69]. Hence, for fabrication simplicity 1 µm is opted to be the optimal disk outer radius.

Similar marginal impact of NiCr Me-NTA disk height variation on emissivity magnitude can be can be observed as shown in Fig. 5(b). NiCr disk height has been varied from 0.05 µm to 0.3 µm to examine the variation effect on emissivity. Again, higher intrinsic ohmic loss influences this trivial behavior. Although, increasing the disk height resulted in slight reduction in emissivity magnitude. Therefore, 0.1 µm of disk height is selected to be the most effective height as it demonstrated highest magnitude of emissivity which is 98% at 4.5 µm.

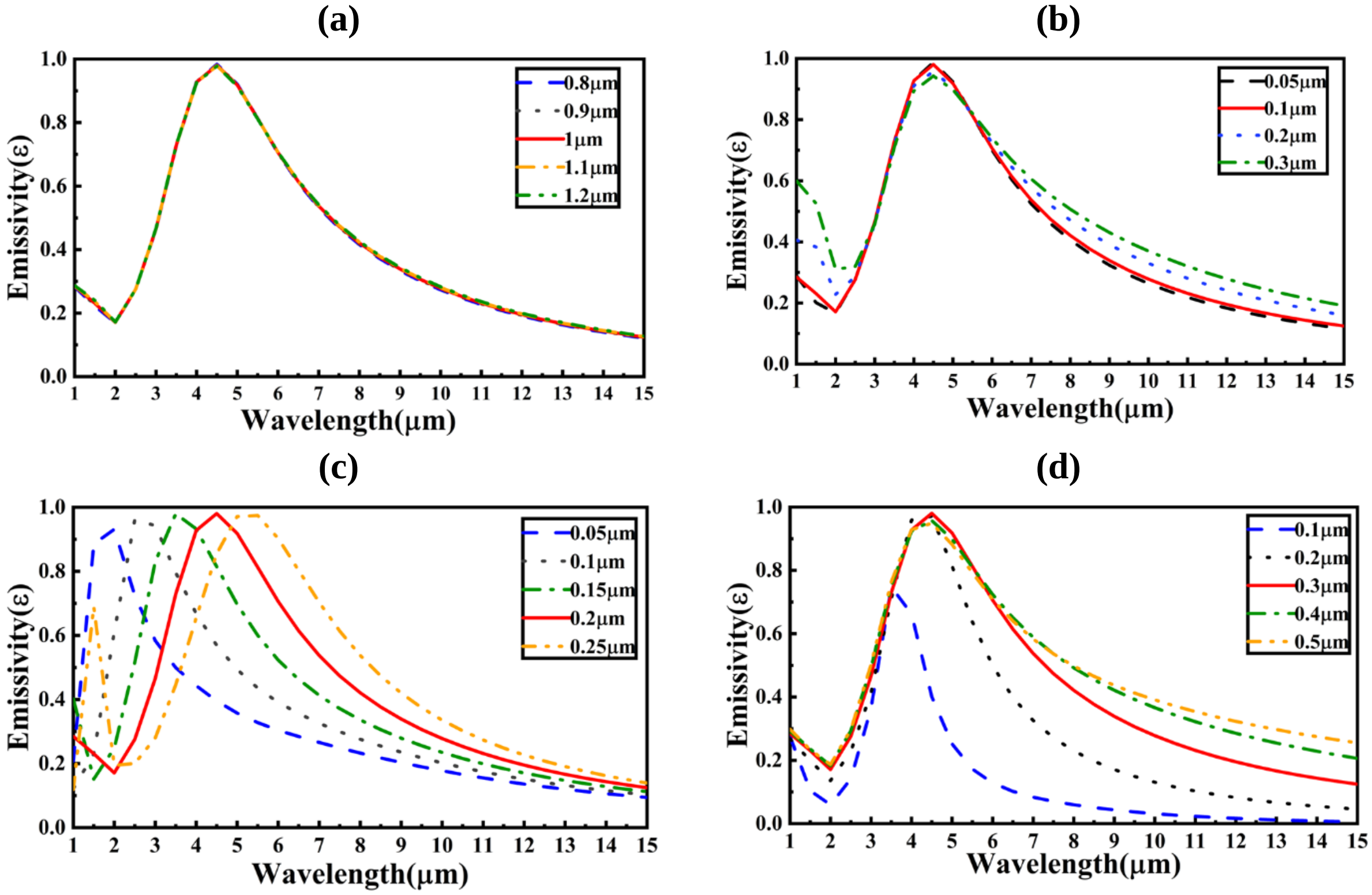

**Fig. 5:** The impact on emissivity spectrum of NiCr Metasurface-based single band IR emitter for variation in (a) disk outer radius, (b) disk height, (c) dielectric spacer thickness, (d) ground plane thickness.

In Fig. 5(c), the effect of SiC dielectric spacer thickness variation on emissivity magnitude at different wavelengths has been illustrated. The thickness of SiC dielectric layer is varied from 0.05 µm to 0.25 µm. As the thickness increases, the emission peak can be observed to be shifting towards longer wavelength which is known as redshift. This trend is transpired because a thicker dielectric layer increases the optical path length between resonator and ground plane. Thus, for larger dielectric spacing, the resonant condition is achieved at longer wavelength [70]. The maximum emission 98% is attained at 4.5 µm for 0.2 µm thickness of SiC dielectric spacer and thus this thickness is chosen to be optimal.

Lastly, the impact of NiCr ground plane thickness alteration on emission magnitude is monitored and picturized in Fig. 5(d). The NiCr ground surface thickness is deviated from 0.1 µm to 0.5 µm. From the figure, it is evident that at lower ground plane thickness, the emissivity is inconsistent and peak magnitude has been reduced. At higher thickness from 0.3 µm to 0.5 µm, the emissivity peak can be seen to be stable and highest magnitude of emission has been achieved at 0.3 µm as this thickness can be considered above the skin depth. Thus, optimal NiCr ground plane thickness is opted to be 0.3 µm.

### 3.2.2 Emissivity spectrum of the NiCr Metasurface-based single band IR emitter

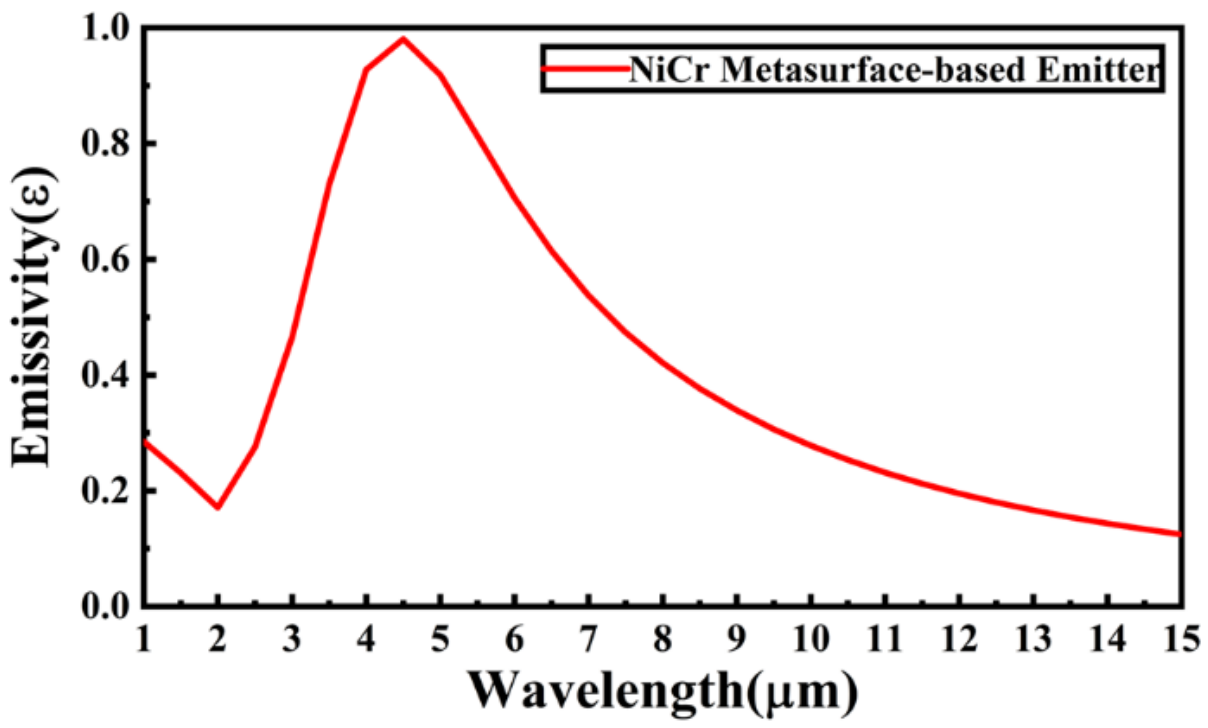

**Fig. 6:** The emissivity spectrum of the proposed NiCr Me-NTA metasurface-based single band IR emitter.

The emissivity spectrum of the NiCr Me-NTA Metasurface-based emitter is depicted in Fig. 6. The wavelength is varied from 1 µm to 15 µm and according emissivity is plotted. It is seen from the figure that near perfect emission is achieved at 4.5 µm and above 90% emission can be observed across 4 µm to 5 µm. Mid-infrared spectrum (MIR) which usually lies between 2.5 µm to 20 µm has many applications in biomedical analysis as many biological molecules shows notable absorption attributes in this wavelength [3]. As a near perfect emission can be seen at the short-wave end of MIR spectrum, it can be evidently stated that this strong emission can assist in more sensitive and selective biomedical measurement.

### 3.2.3 Spectral emittance of the NiCr Metasurface-based single band IR emitter

The emittance spectra obtained by using equation (1) of simulated single band emitter and a reference blackbody with a known temperature are displayed in Fig. 7 (a). The rapid increment in spectral emittance of the emitter results in temperature rise and it has been shown in the figure. When the temperature increases, spectral emittance peak gets shifted towards the short wave direction also emittance band become narrower. At 700 K, it is noticeable that the maximum emittance of our NiCr emitter is 2086.17 W/m$^2$/µm at 4.5 µm wavelength which is nearly that of a perfect blackbody of value is 2127.58 W/m$^2$/µm in that same wavelength. Fig. 7(b) displays that at different temperature; the emitter sample exchanges the radiative heat flux at 3.5-5.5 µm operational band and it has been compared against a blackbody in the same band. Furthermore, the comparison in Fig. 7(b) denotes that at higher temperature 700 K, the exchanged heat flux of a NiCr emitter across 3.5-5.5 µm band with band emissivity 91% is 3698.5 W/m$^2$ which is smaller than that of the blackbody which is 4104.9 W/m$^2$.

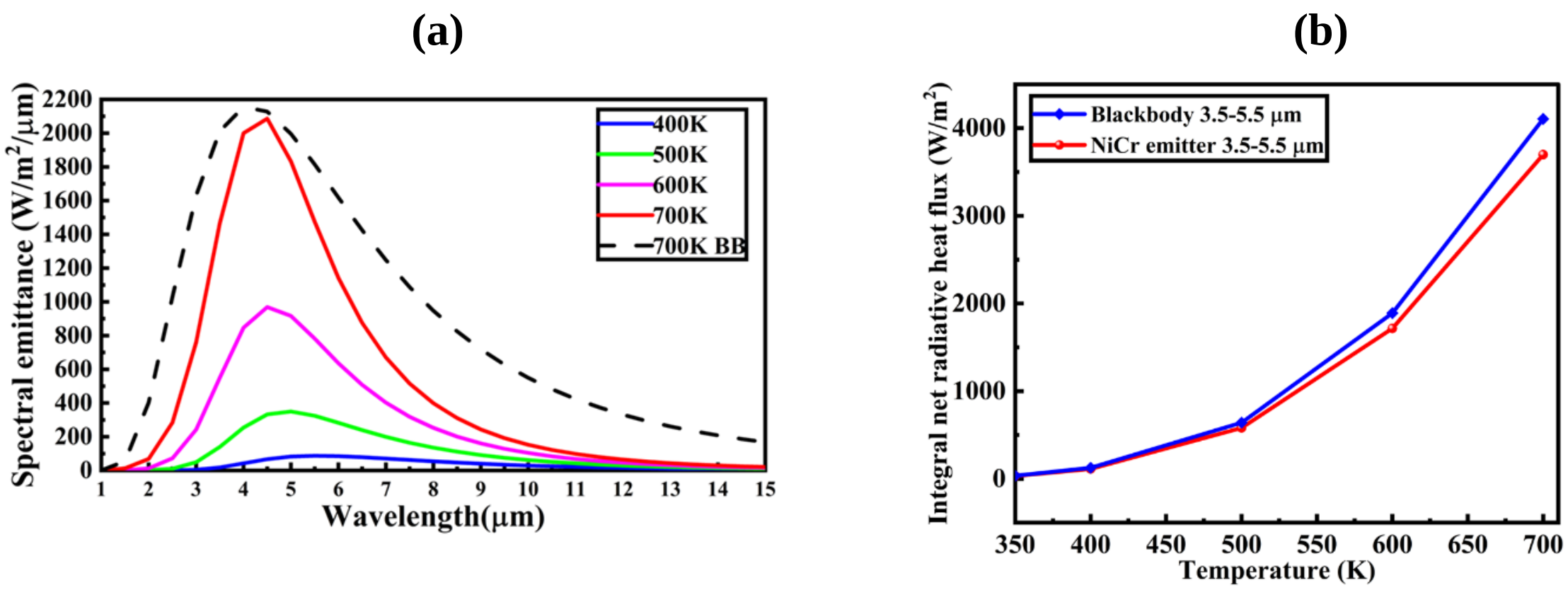

**Fig. 7:** (a) Spectral emittance of the NiCr Me-NTA metasurface-based IR emitter, (b) Net radiative heat flux in 3.5-5.5 µm at various temperature for Black body and the NiCr emitter.

### 3.2.4 Voltage and power analysis of the NiCr Metasurface-based single band IR emitter

The temperature as a function of both voltage and power are illustrated in Fig. 8(a) and (b), respectively. To attain the maximum temperature condition for effective function of NiCr emitter, a 2.23 V electrical DC excitation is required. However, considering emitter as a blackbody altered the requirement of DC excitation to 2.66 V. This highlights that the engineered emission profiles suppress the excess energy consumption, while sustaining strong emissive performance at the desired spectral range. Fig 8(b) demonstrates that the input power required for the NiCr emitter to obtain 700 K is 615 mW and at that time the radiant output power gets 199 mW within that particular band. However, to operate at the same temperature, the blackbody requires 873 mW input power and corresponding radiant power is 221 mW. This validates that the power requirement is reduced to 29.5% at 700 K.

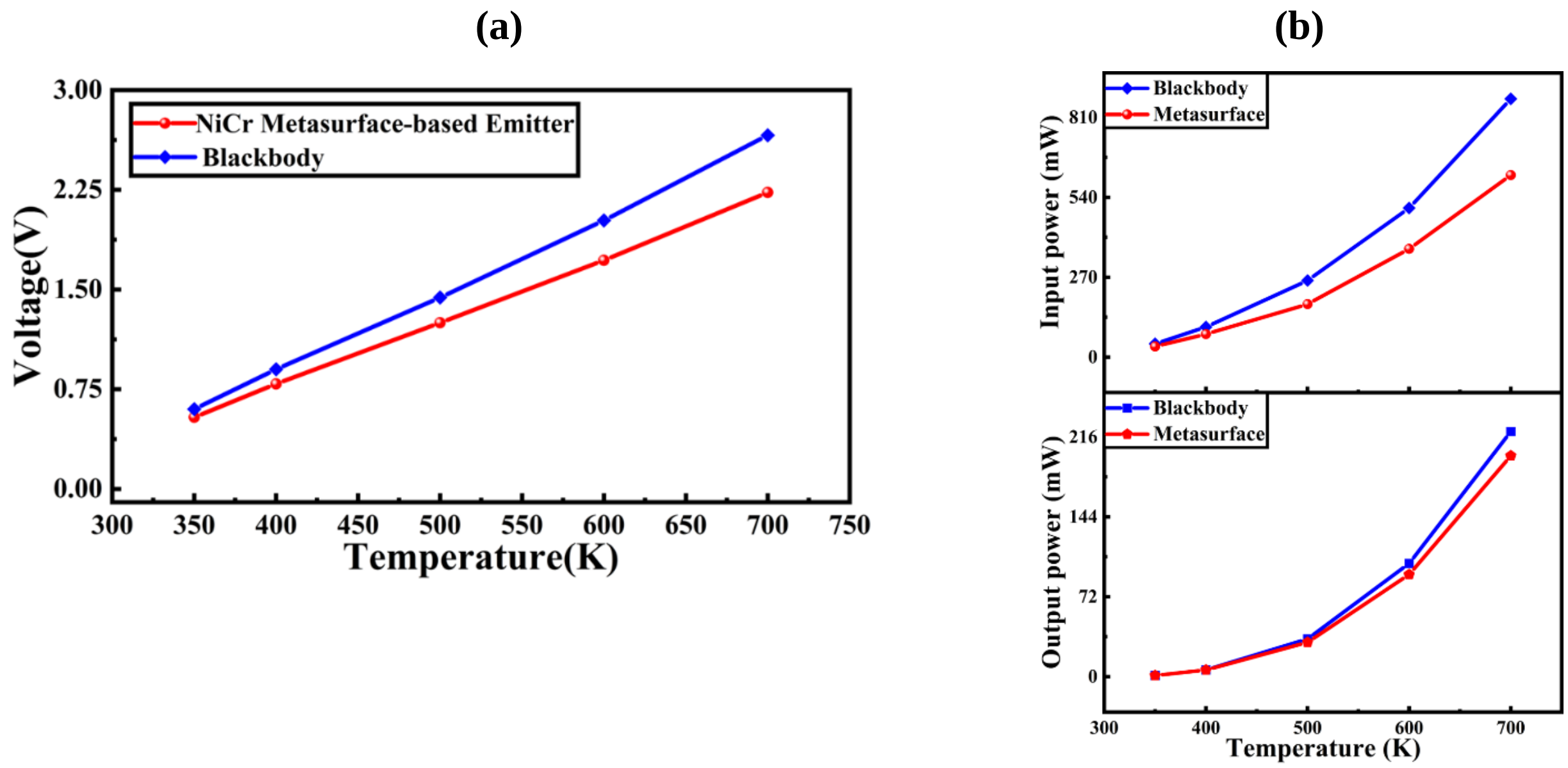

**Fig. 8:** NiCr Me-NTA metasurface-based single band IR emitter (a) voltage analysis and (b) power analysis for different temperature.

### 3.2.5 Efficiency analysis of the NiCr Metasurface-based single band IR emitter

Figure 9 shows that in-band electro-optical conversion efficiency improves with increasing temperature. The proposed NiCr emitter attains a peak efficiency of 32.3% at 700 K, generating 199 mW of in-band radiant power from 615 mW of electrical input power. This trend occurs because thermal radiation grows strongly with temperature therefore, more electrical energy is converted into IR radiation.

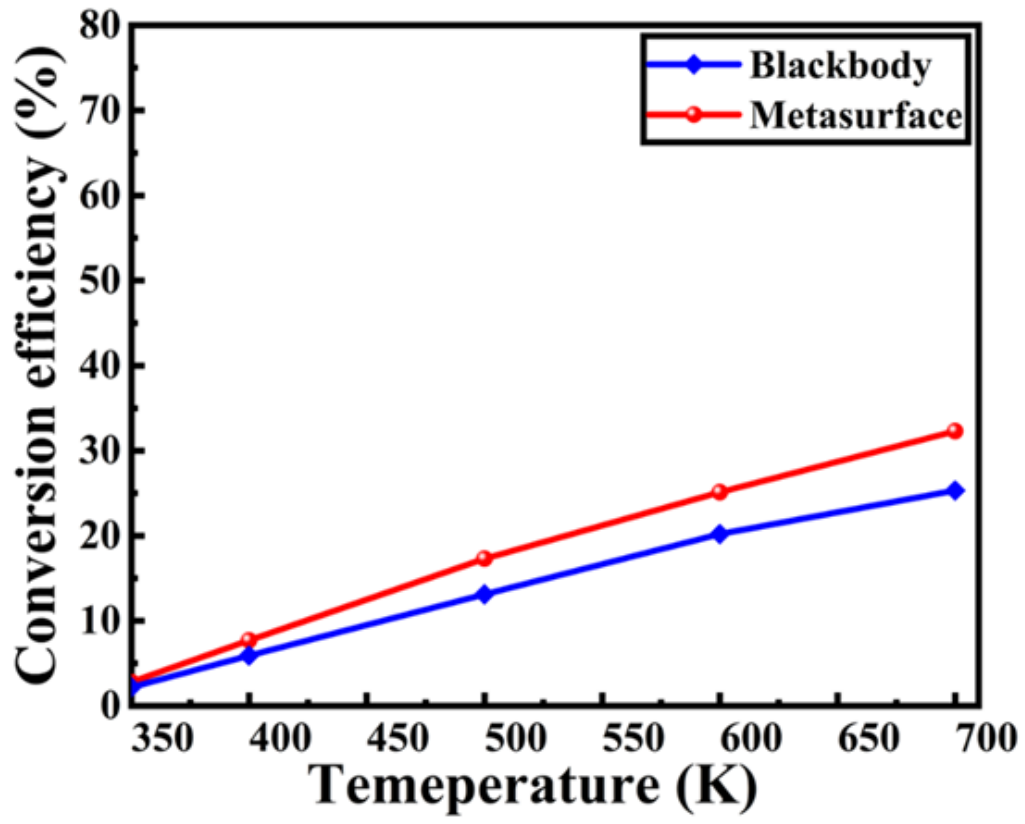

**Fig. 9:** Temperature depended efficiency of the NiCr Me-NTA metasurface-based IR emitter.

#### 3.2.6 Electromagnetic field distribution at resonant wavelength

The electric and magnetic field distributions of NiCr Me-NTA emitter at resonant wavelength are displayed in Fig. 10(a) and (b), respectively to elucidate the emission mechanism. Strong E-field can be observed at the parallel horizontal edges of the ring showed in Fig. 10(a) which indicates the capacitive coupling and results in resonance which is essential for emission. In Fig. 10(b), significant and confined magnetic field distribution is seen around the current-loop regions of the NiCr ring. This presence of E-field and M-field validates the resonance mechanism for the emitter [71].

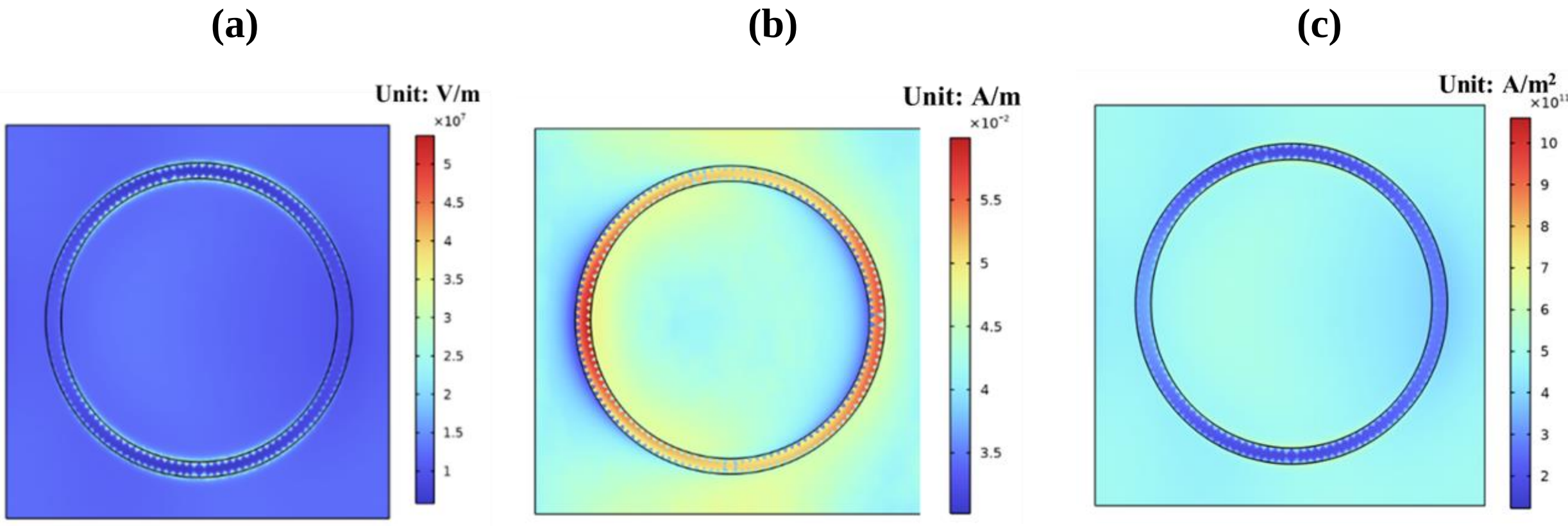

**Fig. 10:** (a) Electric field distribution (b) Magnetic field distribution and (c) Surface current density of NiCr-based IR emitter at 2.5 μm wavelength.

The surface current density of the NiCr Me-NTA-based IR emitter is illustrated in Fig. 10(c). Confined and strong surface current is noticeable along the NiCr ring. Electro-magnetic coupled resonance is validated by these circulating currents and therefore enhanced narrowband emission is achieved through the metasurface.

## 3.3 Parametric and performance analysis of Au Metasurface-based double band IR emitter

### 3.3.1 Parametric analysis of Au Metasurface-based double band IR emitter

Influence of Au based Me-NTA disk outer radius variation on the emissivity spectrum is represented in Fig. 11(a). Disk outer radius has been altered from 0.37 μm to 0.77 μm. While subtle influence at first emission band occurred, significant impact of outer radius variation on second emission band has been observed from the figure which is opposite from that of the first suggested design. Unlike NiCr material which is used in the first device, Au which is used in this structure as metasurface incorporates relatively less intrinsic ohmic loss and hence slight radius deviation results in emission alteration. Moreover, redshift of the band at longer wavelength has been detected with increased disk outer radius. Optimal result is achieved with 0.57 μm disk outer radius where it attained 92.57% emissivity at 2.5 μm and 80.85% is attained at 10 μm and thus this radius has been adopted as the effective Me-NTA disk outer radius for suggested second structure.

Fig. 11(b) represents influence of the Au Me-NTA metasurface disk height alteration on emissivity spectrum. A contrastive impact compared to the disk outer radius has been observed in the illustration. As the disk height increased from 0.05 μm to 0.3 μm, the emission peak can be seen to shift towards shorter wavelength which is called blue shift. This pattern can be explained via modified phase accumulation with increasing disk thickness which blueshifts the resonance [72]. Moreover, 0.1 μm of Au based Me-NTA metasurface disk height is opted to be the most favorable height as it showed maximum magnitude of emissivity of 92.57% at 2.5 μm for first narrowband and 80.85% at 10 μm for second narrowband.

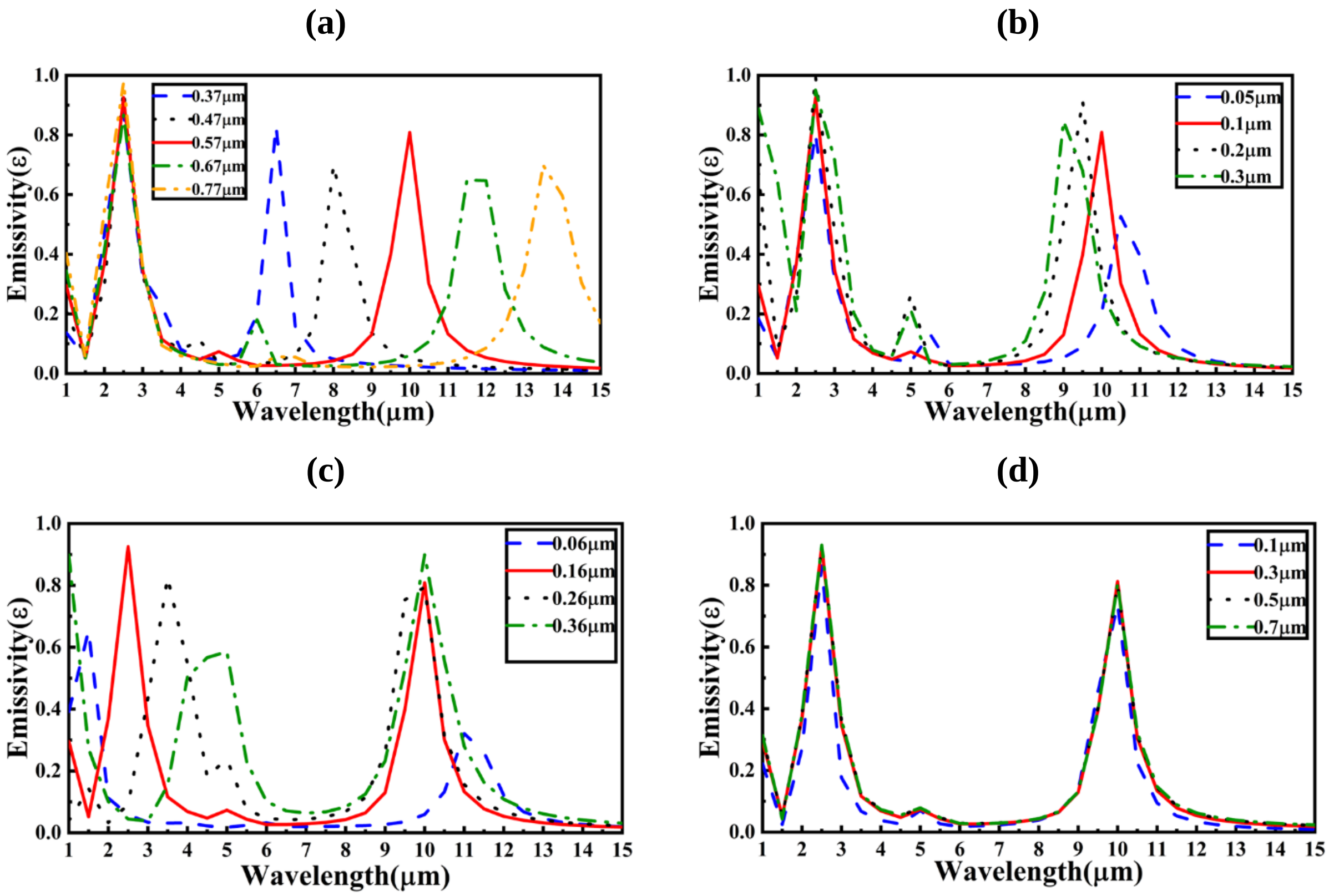

**Fig. 11:** The impact on emissivity spectrum of Au Me-NTA metasurface-based double band IR emitter for variation in (a) disk outer radius, (b) disk height, (c) dielectric spacer thickness, (d) ground plane thickness.

Fig. 11(c) analyses the effect of SiC dielectric spacer thickness variation on emissivity spectrum across 1 µm to 15 µm wavelengths. Here, the dielectric layer thickness is altered from 0.06 µm to 0.36 µm. The graph visualizes that as the thickness elevated, the emission peak of the first band obtained redshift. However, slight shifting of emission peaks with reduced magnitude is detected at the second band. Dielectric thickness of 0.16 µm provided the most effective emission for both band and thereby this thickness is decided to be the optimal thickness for the dielectric spacer of the proposed second structure.

Finally, NiCr ground plane thickness deviation effect is analyzed by evaluating the corresponding emission magnitude and it has been plotted in Fig. 11(d). This ground plane of the suggested second device incorporates 0.01 µm thick layer of Au at top and 0.3 µm thick layer of NiCr at the bottom. For parametric analysis simplicity, only NiCr layer thickness is altered from 0.1 µm to 0.7 μm. Since the overall thickness is larger than the skin depth, the ground plane thickness provides insignificant impact on the emission spectrum. From the graph it is clear that most efficient and stable magnitude of emission has been achieved at 0.3 μm thickness of the ground plane.

### 3.3.2 Emissivity spectra of the Au Metasurface-based double band IR emitter

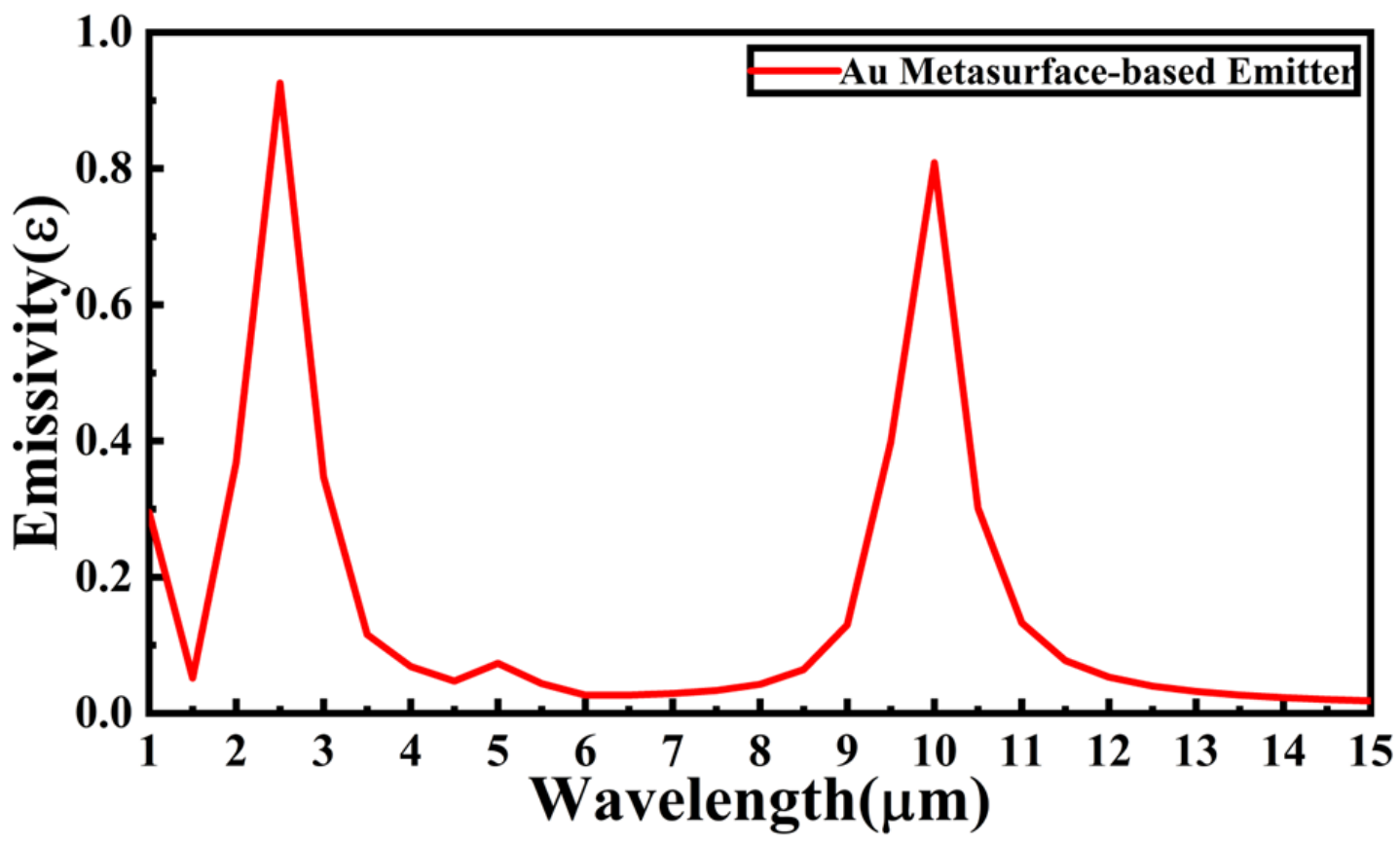

**Fig. 12:** The emissivity spectrum of the proposed Au Me-NTA metasurface-based double band IR emitter.

The emissivity spectrum of the Au metasurface-based emitter is plotted in Fig. 12 and it exhibits two significant emission peaks, one at 2.5 μm and another at 10 μm. First peak displays about 93% emission at 2.5 μm and second peak shows 85% emission at 10 μm. While near perfect emission at 2.5 μm is widely used for biomedical application, strong emission at 10 μm is predominantly used in industrial applications such as thermal imaging, gas sensing etc. Long-wave IR cameras and sensors operate at around 8 μm to 14 μm which can be utilized in industrial thermal imaging systems [73].

### 3.3.3 Spectral emittance of the Au Metasurface-based double band IR emitter

Fig. 13(a) illustrates the spectral emittance of the simulated Au-metasurface based emitter with the reference of the blackbody in two known temperatures. When the temperature increased, the transition from single band to dual band in fig. 13 (a) became very separable and the peak emission occurred at 2.5 μm and 10 μm. The spectral emittance characteristic of the proposed emitter at different temperature dictates both tunability and operational versatility through single and dual band operation. At temperature 500 K, maximum emittance at 10 μm attains spectral emittance of 180.42 W/m$^2$/μm and that of the blackbody is 223.11 W/m$^2$/μm, as observable in the figure. However, at temperature 700 K, it shows a noteworthy maximum emittance at two different wavelengths, one is at 2.5 μm with spectral emittance of 953.19 W/m$^2$/μm and another is at 10 μm with spectral emittance of 444.32 W/m$^2$/μm. For blackbody, the spectral emittances are 1030.16 W/m$^2$/μm and 549.45 W/m$^2$/μm. Furthermore, Fig. 13(a) shows the unnecessary emission got suppressed significantly but near perfect emissions at necessary wavelengths were sustained.

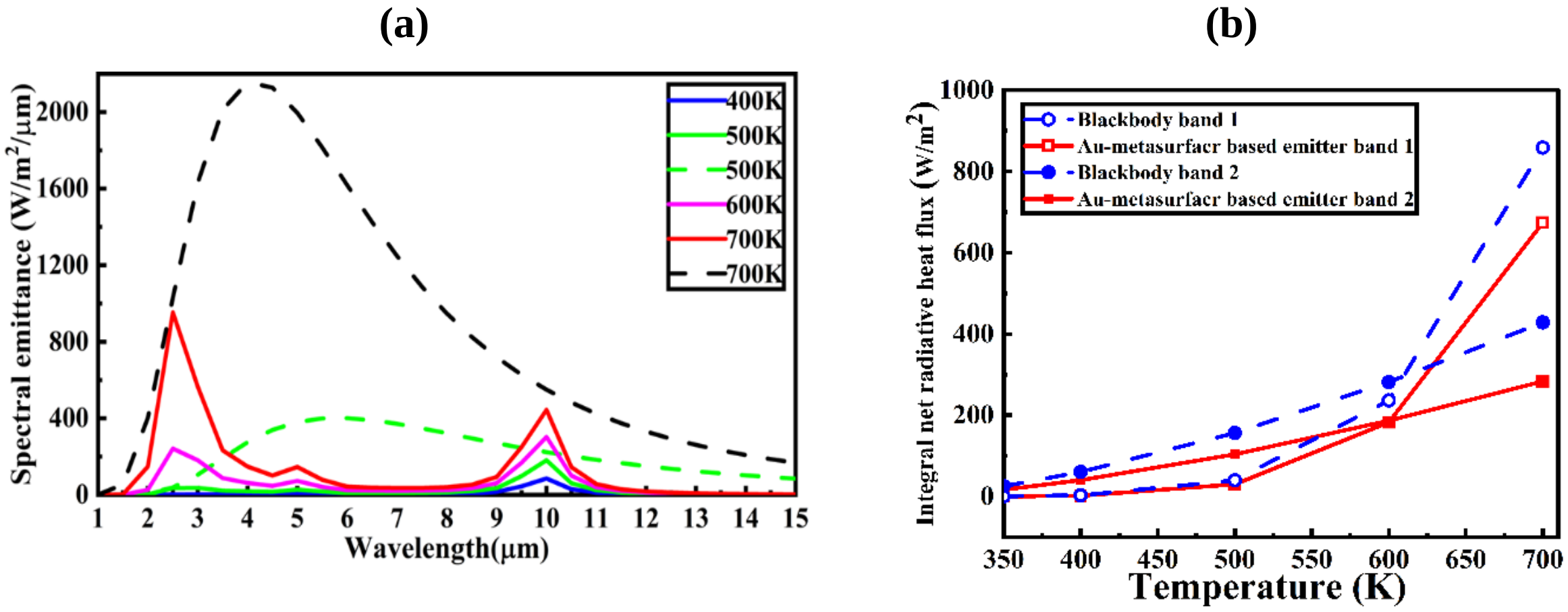

**Fig. 13:** (a) Spectral emittance of the Au Me-NTA metasurface-based IR emitter, (b) Net radiative heat flux in 2.3-3 μm and 9.5-10.3 μm at various temperature for Black body and the Au Me-NTA emitter.

The net radiative heat flux exchange in two operational bands 2.3-3 μm and 9.5-10.3 μm at various temperature and parallelism with the blackbody has been described in Fig. 13(b). When temperature is 500 K, the net radiative heat flux is 103.86 W/m$^2$ in 9.5-10.3 μm band with the band emissivity of 66%. At higher temperature 700 K, band 1 with band emissivity 78% has 673.71 W/m$^2$ and 283.43 W/m$^2$ in band 2 with emissivity 66%.

### 3.3.4 Voltage and power analysis of the Au Metasurface-based double band IR emitter

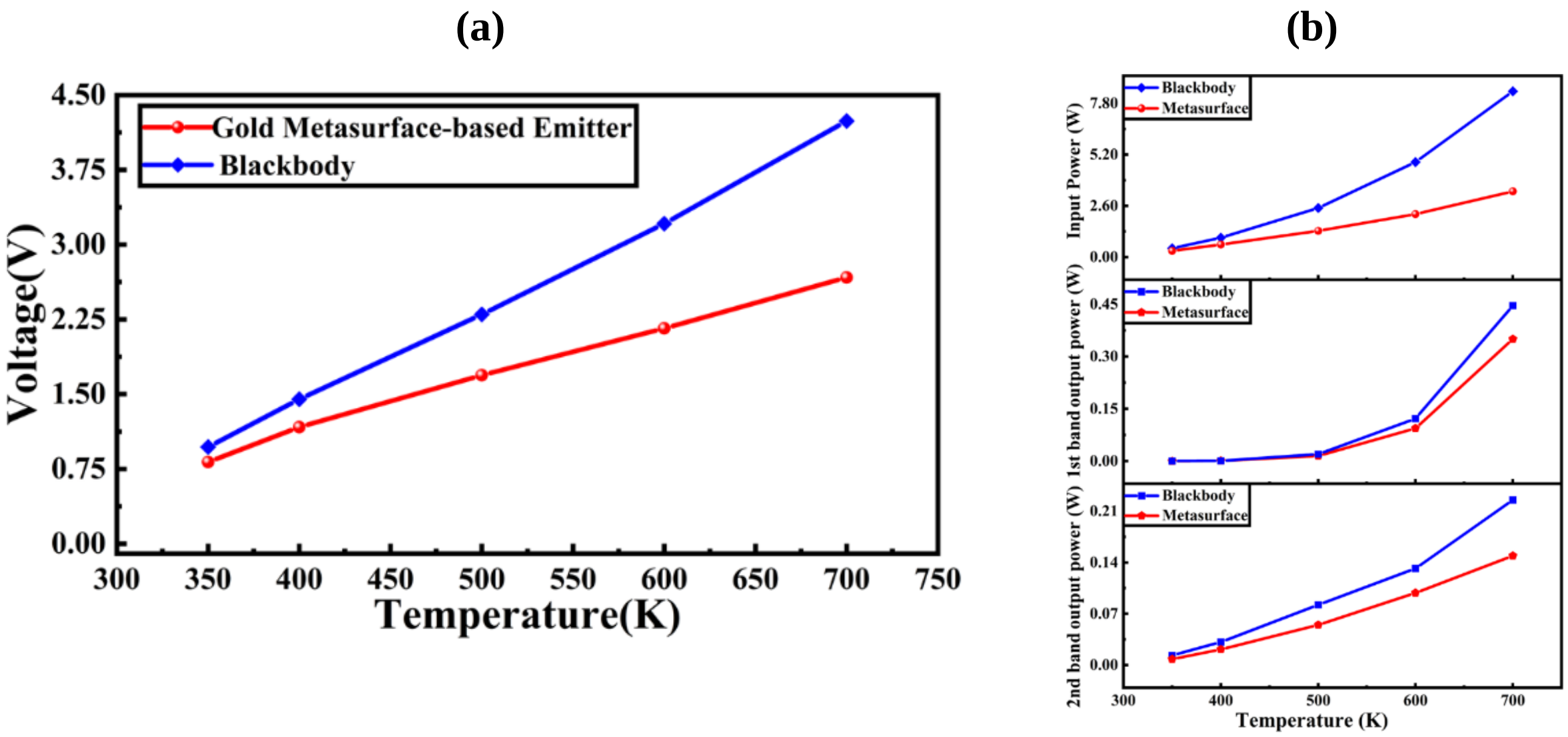

**Fig. 14:** Au Au Me-NTA metasurface-based double band IR emitter (a) voltage analysis and (b) power analysis for different temperature.

Fig. 14(a) depicts that for Au metasurface-based emitter, to reach temperature 500 K and 700 K, it requires voltages 1.69 V and 2.67 V and that for blackbody are 2.3 V and 4.24 V, accordingly. Fig 14(b) shows that to achieve 500 K and 700 K, Au metasurface-based emitter requires power of 1.33 W and 3.33 W and that for blackbody its 2.49 W and 8.41 W, correspondingly. When temperature is 500 K the output radiant power at band 2 is 0.054 W for Au-emitter and 0.081 W for blackbody. At maximum temperature 700 K, band 1 and band 2 radiant output are 0.35 W and 0.147 W for Au-emitter and 0.446 W and 0.222 W for blackbody. The inspection of Fig. 14(a) and (b) clearly substantiates the context of high-power emission. At 500 K, band 2 is dominant and at 700 K, both bands are dominant.

### 3.3.5 Efficiency analysis of the Au Metasurface-based double band IR emitter

The maximum efficiency of Au Metasurface-based is 10.4% for band 1 and 4.4% for band 2 at 700K as shown in Fig. 15. However, at 500K band 2 has efficiency of 4.6%. The increasing conversion efficiency of band 1 with temperature and decreasing conversion efficiency in band 2 appears because of temperature dependent spectral shift of thermal radiation. In accordance with Planck's law and Wien's displacement law, rising temperature shifts the emission spectrum toward shorter wavelength direction resulting in radiation enhancement for band 1 consequently increasing CE. However, this reduces relative spectral contribution around band 2 leading to CE decrement for this band.

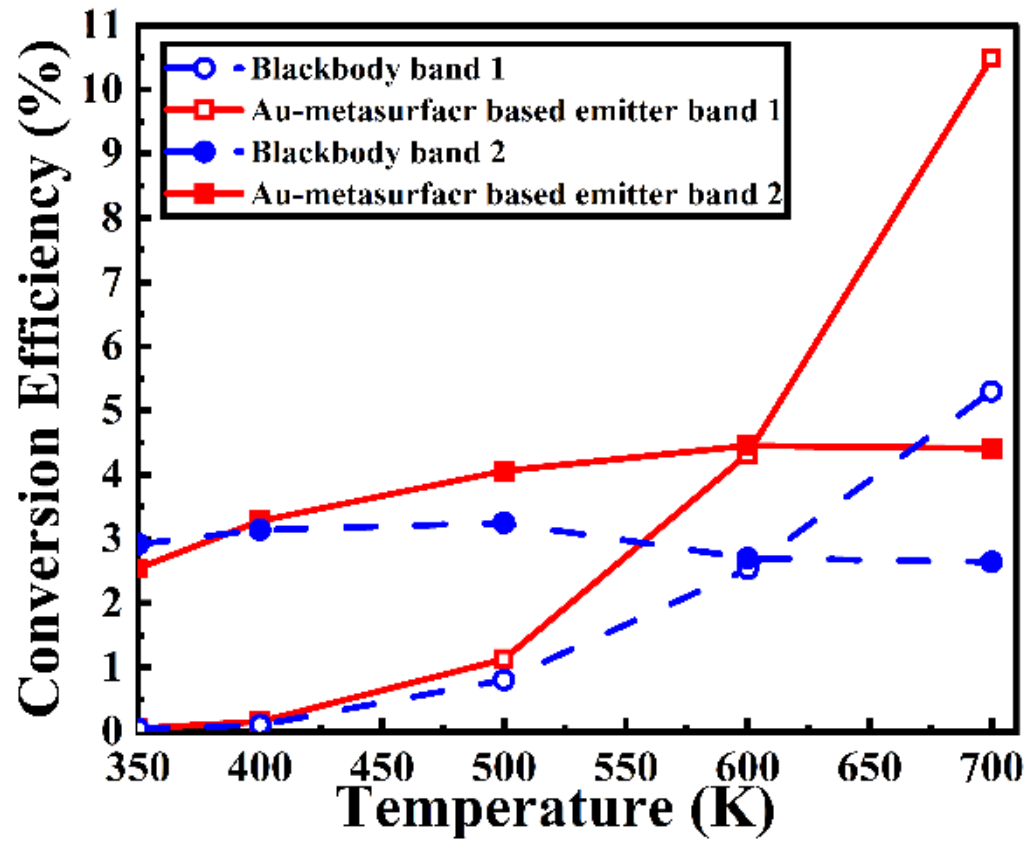

**Fig. 15:** Temperature depended efficiency of the Au Au Me-NTA metasurface-based IR emitter.

### 3.3.6 Electromagnetic field distribution and surface current density at resonant wavelengths

E-field distribution for Au Me-NTA -based IR emitter at both resonant wavelengths namely 2.5 μm and 10 μm are plotted in Fig. 16(a) and (b), respectively. In Fig. 16(a) for 2.5 μm, confined E-field is observed along Au ring circumference and substantial E-field can be seen at the inner edge of the ring. This confirms profound emission at this wavelength. On the other hand, in Fig. 16(b) for 10 μm, localized E-field has been noticed however comparatively throughout smaller area than at 2.5 μm which validates its slightly lesser magnitude of emission. In Fig. 16(c) and (d), the magnetic field distribution for Au-based IR emitter is illustrated. At 2.5 μm showed in Fig. 16(c), substantial M-field is spotted across spread area of the surface which signifies narrower emission but with higher magnitude. On the contrary, at 10 μm in Fig. 16(d), higher magnitude of M-field is noticed at the vertical ring edge while total distribution area is comparatively lesser than that at 2.5 μm, generating relatively lesser emission at this wavelength.

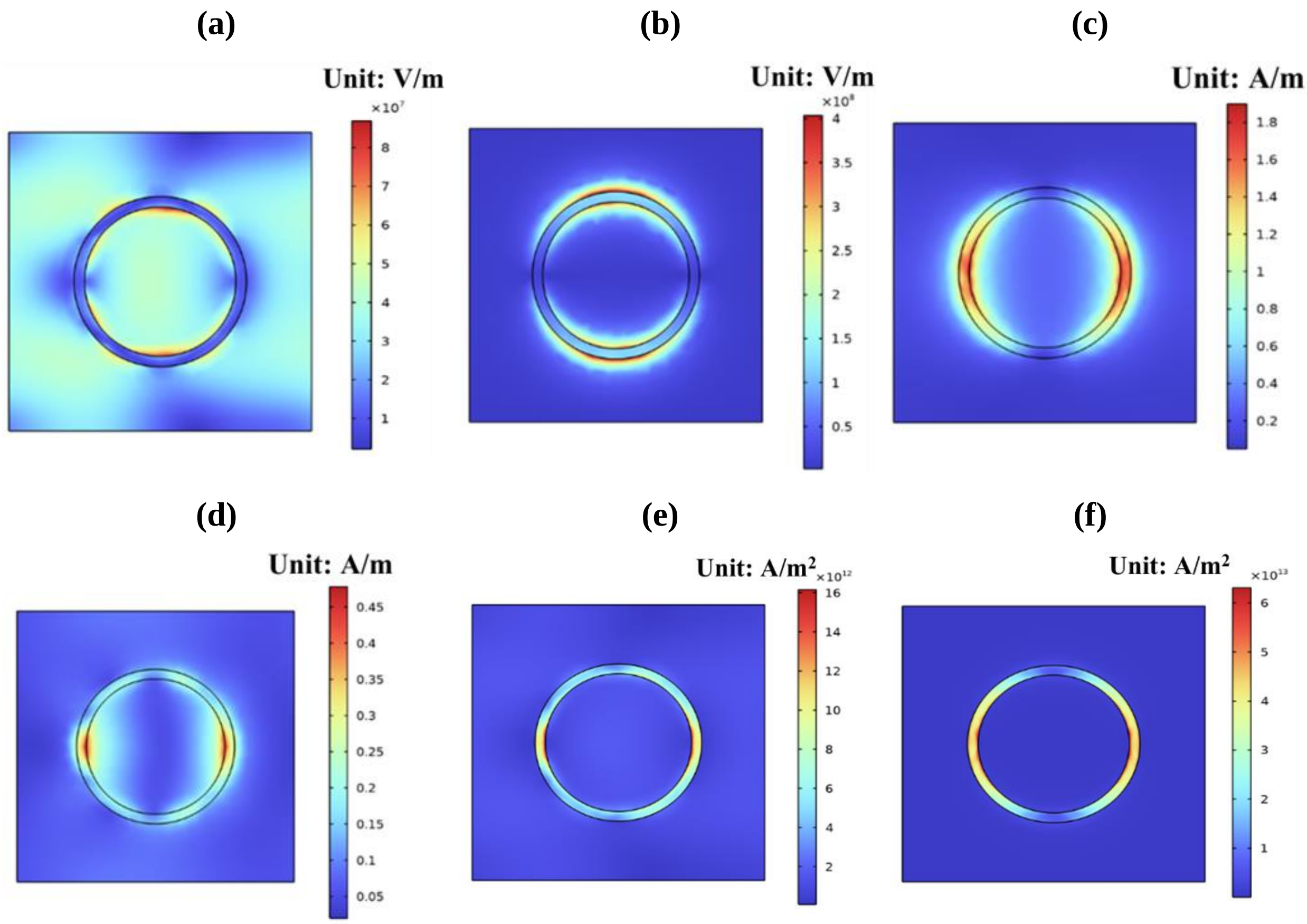

**Fig. 16:** Electromagnetic field distributions of the proposed Au-based IR emitter: (a,b) electric field, (c,d) magnetic field, and (e,f) surface current density at wavelengths of 2.5 μm and 10 μm, respectively.

Surface current density of Au-based IR emitter at resonance wavelengths are plotted in Fig. 16(e) and (f). For 2.5 μm in Fig. 16(e), moderate to high surface current is localized across the Au ring perimeter. This surface current realizes the settings for resonance which as a result generates emission at the metasurface. For 10 μm in Fig. 16(f), similar surface current density is observed but throughout lesser area compared to the previously mentioned wavelength. This justifies lesser magnitude of emission at 10 μm than 2.5 μm.

## 4. Conclusion

Two high power IR emitters, namely, NiCr Me-NTA -metasurface based emitter that emits single narrowband at MIR by NiCr heater and Au Me-NTA-metasurface based emitter that results in double narrowband emission at SIR and FIR by NiCr-Au heater is proposed. Both heaters are based on DC bias voltage which is later converted into high heat energy with high temperature uniformity. The FEM simulation results provide insight into the emissivity and later this manifested as a significant alteration of the blackbody spectrum. Both the heaters archive high in-band conversion efficiency (CE) in 700K which requires 2.23 V for the NiCr-based heater and 2.67 V for NiCr-Au based heater. For the dual band emitter, only FIR emission with less consumable power is demonstrated at 500K for 1.69V. By introducing meta surface on both emitters, the wavelength filtering effect has been obtained that results in single-band emission with center wavelength at 4.5 μm for the NiCr Me-NTA -metasurface based emitter and double-band emission with center wavelength at 2.5 μm and 10 μm for the Au Me-NTA -metasurface based emitter. The highest CE for the single-band emitter in desired band is 32.3% with a radiated power of 199mW. Likewise, for the dual-band emitter operating at 700K, highest achieves CE of 10.4% in first band and 4.4% in second band with radiated power 350mW and 147mW, accordingly. These findings, expands the possibilities of our proposed IR emitters in numerous applications in both industrial and biomedical field.

*Corresponding Author: E-mail: jak_apee@ru.ac.bd (Jaker Hossain).

**Disclosure:** None of the authors have any competing financial interest.

**Data availability:** Data will be available from the corresponding author upon reasonable request.

**Declaration of generative AI and AI-assisted technologies:** None of the authors use any AI or AI-assisted technologies in writing this manuscript.

## References

[1] M.A. Butt, M. Juchniewicz, M. Słowikowski, Ł. Kozłowski, R. Piramidowicz, Mid-Infrared Photonic Sensors: Exploring Fundamentals, Advanced Materials, and Cutting-Edge Applications, Sensors 25 (2025) 1102. https://doi.org/10.3390/s25041102.

[2] M. Hlavatsch, B. Mizaikoff, Advanced mid-infrared lightsources above and beyond lasers and their analytical utility, Analytical Sciences 38 (2022) 1125–1139. https://doi.org/10.1007/s44211-022-00133-3.

[3] T. Yin, T. Yang, X. Dong, B. Liu, Z. Qiao, Mid-infrared spectroscopy and its applications in biomedical analysis and imaging, Biochem. Biophys. Res. Commun. 790 (2025) 152890. https://doi.org/10.1016/j.bbrc.2025.152890.

[4] T. Ott, M. Schossig, V. Norkus, G. Gerlach, Efficient thermal infrared emitter with high radiant power, Journal of Sensors and Sensor Systems 4 (2015) 313–319. https://doi.org/10.5194/jsss-4-313-2015.

[5] Y. Hamada, F. Teraoka, T. Matsumoto, A. Madachi, F. Toki, E. Uda, R. Hase, J. Takahashi, N. Matsuura, Effects of far infrared ray on Hela cells and WI-38 cells, Int. Congr. Ser. 1255 (2003) 339–341. https://doi.org/10.1016/S0531-5131(03)00887-2.

[6] T.-V. Dinh, I.-Y. Choi, Y.-S. Son, J.-C. Kim, A review on non-dispersive infrared gas sensors: Improvement of sensor detection limit and interference correction, Sens. Actuators B Chem. 231 (2016) 529–538. https://doi.org/10.1016/j.snb.2016.03.040.

[7] Y. Gong, Z. Wang, K. Li, L. Uggalla, J. Huang, N. Copner, Y. Zhou, D. Qiao, J. Zhu, Highly efficient and broadband mid-infrared metamaterial thermal emitter for optical gas sensing, Opt. Lett. 42 (2017) 4537. https://doi.org/10.1364/OL.42.004537.

[8] L.L. Rajeswara Rao, M.K. Singha, K.M. Subramaniam, N. Jampana, S. Asokan, Molybdenum Microheaters for MEMS-Based Gas Sensor Applications: Fabrication, Electro-Thermo-Mechanical and Response Characterization, IEEE Sens. J. 17 (2017) 22–29. https://doi.org/10.1109/JSEN.2016.2621179.

[9] R. Xu, Y.-S. Lin, Tunable Infrared Metamaterial Emitter for Gas Sensing Application, Nanomaterials 10 (2020) 1442. https://doi.org/10.3390/nano10081442.

[10] O.S. Çifçi, Narrow Bandwidth and Tunable Mid-Infrared Thermal Emitter Design Based on Double Asymmetric Dielectric Metasurfaces, International Journal of Advances in Engineering and Pure Sciences 36 (2024) 320–325. https://doi.org/10.7240/jeps.1529681.

[11] K. Sun, Y. Cai, L. Huang, Z. Han, Ultra-narrowband and rainbow-free mid-infrared thermal emitters enabled by a flat band design in distorted photonic lattices, Nat. Commun. 15 (2024) 4019. https://doi.org/10.1038/s41467-024-48499-4.

[12] Y. Gong, K. Li, N. Copner, H. Liu, M. Zhao, B. Zhang, A. Pusch, D.L. Huffaker, S.S. Oh, Integrated and spectrally selective thermal emitters enabled by layered metamaterials, Nanophotonics 10 (2021) 1285–1293. https://doi.org/10.1515/nanoph-2020-0578.

[13] K. Sun, Y. Cai, L. Huang, Z. Han, Ultra-narrowband and rainbow-free mid-infrared thermal emitters enabled by a flat band design in distorted photonic lattices, Nat. Commun. 15 (2024) 4019. https://doi.org/10.1038/s41467-024-48499-4.

[14] X. Liu, T. Tyler, T. Starr, A.F. Starr, N.M. Jokerst, W.J. Padilla, Taming the Blackbody with Infrared Metamaterials as Selective Thermal Emitters, Phys. Rev. Lett. 107 (2011) 045901. https://doi.org/10.1103/PhysRevLett.107.045901.

[15] H. Ou, F. Lu, Z. Xu, Y.-S. Lin, Terahertz Metamaterial with Multiple Resonances for Biosensing Application, Nanomaterials 10 (2020) 1038. https://doi.org/10.3390/nano10061038.

[16] W. Huang, R. Xu, Y.-S. Lin, C.-H. Chen, Three-dimensional pyramid metamaterial with tunable broad absorption bandwidth, AIP Adv. 10 (2020). https://doi.org/10.1063/1.5142440.

[17] Z. Liang, Y. Wen, Z. Zhang, Z. Liang, Z. Xu, Y.-S. Lin, Plasmonic metamaterial using metal-insulator-metal nanogratings for high-sensitive refraction index sensor, Results Phys. 15 (2019) 102602. https://doi.org/10.1016/j.rinp.2019.102602.

[18] J. Luo, Y.-S. Lin, High-efficiency of infrared absorption by using composited metamaterial nanotubes, Appl. Phys. Lett. 114 (2019). https://doi.org/10.1063/1.5063736.

[19] Z. Xu, R. Xu, J. Sha, B. Zhang, Y. Tong, Y.-S. Lin, Infrared metamaterial absorber by using chalcogenide glass material with a cyclic ring-disk structure, OSA Contin. 1 (2018) 573. https://doi.org/10.1364/OSAC.1.000573.

[20] R. Xu, Y.-S. Lin, Characterizations of reconfigurable infrared metamaterial absorbers, Opt. Lett. 43 (2018) 4783. https://doi.org/10.1364/OL.43.004783.

[21] R. Xu, J. Luo, J. Sha, J. Zhong, Z. Xu, Y. Tong, Y.-S. Lin, Stretchable IR metamaterial with ultra-narrowband perfect absorption, Appl. Phys. Lett. 113 (2018). https://doi.org/10.1063/1.5044225.

[22] Y.-S. Lin, W. Chen, Perfect meta-absorber by using pod-like nanostructures with ultra-broadband, omnidirectional, and polarization-independent characteristics, Sci. Rep. 8 (2018) 7150. https://doi.org/10.1038/s41598-018-25728-7.

[23] N.I. Landy, S. Sajuyigbe, J.J. Mock, D.R. Smith, W.J. Padilla, Perfect Metamaterial Absorber, Phys. Rev. Lett. 100 (2008) 207402. https://doi.org/10.1103/PhysRevLett.100.207402.

[24] C.M. Soukoulis, M. Wegener, Past achievements and future challenges in the development of three-dimensional photonic metamaterials, Nat. Photonics 5 (2011) 523–530. https://doi.org/10.1038/nphoton.2011.154.

[25] H.-T. Chen, Interference theory of metamaterial perfect absorbers, Opt. Express 20 (2012) 7165. https://doi.org/10.1364/OE.20.007165.

[26] R. Alaee, M. Farhat, C. Rockstuhl, F. Lederer, A perfect absorber made of a graphene micro-ribbon metamaterial, Opt. Express 20 (2012) 28017. https://doi.org/10.1364/OE.20.028017.

[27] W. Li, J. Valentine, Metamaterial Perfect Absorber Based Hot Electron Photodetection, Nano Lett. 14 (2014) 3510–3514. https://doi.org/10.1021/nl501090w.

[28] Y. Ra'di, C.R. Simovski, S.A. Tretyakov, Thin Perfect Absorbers for Electromagnetic Waves: Theory, Design, and Realizations, Phys. Rev. Appl. 3 (2015) 037001. https://doi.org/10.1103/PhysRevApplied.3.037001.

[29] J.W. Park, P. Van Tuong, J.Y. Rhee, K.W. Kim, W.H. Jang, E.H. Choi, L.Y. Chen, Y. Lee, Multi-band metamaterial absorber based on the arrangement of donut-type resonators, Opt. Express 21 (2013) 9691. https://doi.org/10.1364/OE.21.009691.

[30] C.-M. Wang, D.-Y. Feng, Omnidirectional thermal emitter based on plasmonic nanoantenna arrays, Opt. Express 22 (2014) 1313. https://doi.org/10.1364/OE.22.001313.

[31] Y.-S. Lin, J. Dai, Z. Zeng, B.-R. Yang, Metasurface Color Filters Using Aluminum and Lithium Niobate Configurations, Nanoscale Res. Lett. 15 (2020) 77. https://doi.org/10.1186/s11671-020-03310-3.

[32] Y.-S. Lin, W. Chen, A large-area, wide-incident-angle, and polarization-independent plasmonic color filter for glucose sensing, Opt. Mater. (Amst). 75 (2018) 739–743. https://doi.org/10.1016/j.optmat.2017.11.043.

[33] Y. Wen, Z. Liang, Y.-S. Lin, C.-H. Chen, Active modulation of polarization-sensitive infrared metamaterial, Opt. Commun. 463 (2020) 125489. https://doi.org/10.1016/j.optcom.2020.125489.

[34] F. Zhan, Y.-S. Lin, Tunable multiresonance using complementary circular metamaterial, Opt. Lett. 45 (2020) 3633. https://doi.org/10.1364/OL.394137.

[35] T. Xu, Y.-S. Lin, Tunable Terahertz Metamaterial Using an Electric Split-Ring Resonator with Polarization-Sensitive Characteristic, Applied Sciences 10 (2020) 4660. https://doi.org/10.3390/app10134660.

[36] X. Chen, Y.-S. Lin, Polarization-Sensitive Metamaterials with Tunable Multi-Resonance in the Terahertz Frequency Range, Crystals (Basel). 10 (2020) 611. https://doi.org/10.3390/cryst10070611.

[37] W. Yang, Y.-S. Lin, Tunable metamaterial filter for optical communication in the terahertz frequency range, Opt. Express 28 (2020) 17620. https://doi.org/10.1364/OE.396620.

[38] Y.J. Kim, Y.J. Yoo, K.W. Kim, J.Y. Rhee, Y.H. Kim, Y. Lee, Dual broadband metamaterial absorber, Opt. Express 23 (2015) 3861. https://doi.org/10.1364/OE.23.003861.

[39] R. Xu, Y.-S. Lin, Tunable Infrared Metamaterial Emitter for Gas Sensing Application, Nanomaterials 10 (2020) 1442. https://doi.org/10.3390/nano10081442.

[40] H.K. Njoto, A.K. Altama, W.-X. Lu, T.-H. Chang, K.-C. Lin, S.-L. Lee, J.P. Chu, C.-T. Lin, Metallic Nanotube Array Enabling Far Infrared Thermal Emitter Performance Enhancement, IEEE Photonics J. 16 (2024) 1–6. https://doi.org/10.1109/JPHOT.2024.3488120.

[41] J.-J. Greffet, R. Carminati, K. Joulain, J.-P. Mulet, S. Mainguy, Y. Chen, Coherent emission of light by thermal sources, Nature 416 (2002) 61–64. https://doi.org/10.1038/416061a.

[42] T. Inoue, M. De Zoysa, T. Asano, S. Noda, Realization of narrowband thermal emission with optical nanostructures, Optica 2 (2015) 27. https://doi.org/10.1364/OPTICA.2.000027.

[43] B. Xuan Khuyen, N. Van Ngoc, D. Ngoc Dung, N.P. Hai, N.T. Tung, B. Son Tung, V. Dinh Lam, H. Truong Giang, P.D. Tan, L. Chen, H. Zheng, Y. Lee, Dual-band infrared metamaterial perfect absorber for narrow-band thermal emitters, J. Phys. D Appl. Phys. 57 (2024) 285501. https://doi.org/10.1088/1361-6463/ad3bc7.

[44] D. Popa, R. Hopper, S.Z. Ali, M.T. Cole, Y. Fan, V.-P. Veigang-Radulescu, R. Chikkaraddy, J. Nallala, Y. Xing, J. Alexander-Webber, S. Hofmann, A. De Luca, J.W. Gardner, F. Udrea, A highly stable, nanotube-enhanced, CMOS-MEMS thermal emitter for mid-IR gas sensing, Sci. Rep. 11 (2021) 22915. https://doi.org/10.1038/s41598-021-02121-5.

[45] A. Lochbaum, Y. Fedoryshyn, A. Dorodnyy, U. Koch, C. Hafner, J. Leuthold, On-Chip Narrowband Thermal Emitter for Mid-IR Optical Gas Sensing, ACS Photonics 4 (2017) 1371–1380. https://doi.org/10.1021/acsphotonics.6b01025.

[46] C. Gong, G. Zheng, Selective Properties of Mid-Infrared Tamm Phonon-Polaritons Emitter with Silicon Carbide-Based Structures, Micromachines (Basel). 13 (2022) 920. https://doi.org/10.3390/mi13060920.

[47] T. Jin, J. Zhou, P.T. Lin, Real-time and non-destructive hydrocarbon gas sensing using mid-infrared integrated photonic circuits, RSC Adv. 10 (2020) 7452–7459. https://doi.org/10.1039/C9RA10058J.

[48] D. Popa, F. Udrea, Towards Integrated Mid-Infrared Gas Sensors, Sensors 19 (2019) 2076. https://doi.org/10.3390/s19092076.

[49] S.K. Kamrava, M. Behtaj, Y. Ghavami, S. Shahabi, M. Jalessi, E.E. Afshar, S. Maleki, Evaluation of diagnostic values of photodynamic diagnosis in identifying the dermal and mucosal squamous cell carcinoma, Photodiagnosis Photodyn. Ther. 9 (2012) 293–298. https://doi.org/10.1016/j.pdpdt.2012.03.004.

[50] S. Das, J. Akhtar, Comparative Study on Temperature Coefficient of Resistance (TCR) of the E-beam and Sputter Deposited Nichrome Thin Film for Precise Temperature Control of Microheater for MEMS Gas Sensor, in: 2014: pp. 495–497. https://doi.org/10.1007/978-3-319-03002-9_124.

[51] M.K. Sinha, S.K. Mukherjee, B. Pathak, R.K. Paul, P.K. Barhai, Effect of deposition process parameters on resistivity of metal and alloy films deposited using anodic vacuum arc technique, Thin Solid Films 515 (2006) 1753–1757. https://doi.org/10.1016/j.tsf.2006.06.028.

[52] D. Walter, A. Bülau, S. Bengsch, K. Gläser, A. Zimmermann, Fabrication and Characterization of a Thermal Flow Sensor Based on the Ensinger Microsystems Technology, Metrology 5 (2025) 41. https://doi.org/10.3390/metrology5030041.

[53] Design and Synthesis of Neutral Density Filter in Visible Range for Color Sorting Applications, Indian Journal of Pure & Applied Physics (2024). https://doi.org/10.56042/ijpap.v62i10.8277.

[54] Q. Feng, M. Pu, C. Hu, X. Luo, Engineering the dispersion of metamaterial surface for broadband infrared absorption, Opt. Lett. 37 (2012) 2133. https://doi.org/10.1364/OL.37.002133.

[55] S. Erkan, E. Arpat, S. Peters, Investigation of percolation thickness of sputter coated thin NiCr films on clear float glass, Appl. Surf. Sci. 421 (2017) 373–377. https://doi.org/10.1016/j.apsusc.2017.01.143.

[56] V.B. Svetovoy, P.J. van Zwol, G. Palasantzas, J.Th.M. De Hosson, Optical properties of gold films and the Casimir force, Phys. Rev. B 77 (2008) 035439. https://doi.org/10.1103/PhysRevB.77.035439.

[57] Md.N. Hasan, D. Acharjee, D. Kumar, A. Kumar, S. Maity, Simulation of Low Power Heater for Gas Sensing Application, Procedia Comput. Sci. 92 (2016) 213–221. https://doi.org/10.1016/j.procs.2016.07.348.

[58] M. Qian, Q. Shi, L. Qin, J. Huang, C. Guo, Y. Liu, K. Yu, Fabrication of Selective Thermal Emitter with Multilayer Films for Mid-/Low-Temperature Infrared Stealth with Radiative Cooling, Photonics 10 (2023) 645. https://doi.org/10.3390/photonics10060645.

[59] T.R. Steiner, High temperature steady-state experiment for computational radiative heat transfer validation using COMSOL and ANSYS, Results in Engineering 13 (2022) 100354. https://doi.org/10.1016/j.rineng.2022.100354.

[60] M.K. Sinha, S.K. Mukherjee, B. Pathak, R.K. Paul, P.K. Barhai, Effect of deposition process parameters on resistivity of metal and alloy films deposited using anodic vacuum arc technique, Thin Solid Films 515 (2006) 1753–1757. https://doi.org/10.1016/j.tsf.2006.06.028.

[61] ® ® CHEMICAL COMPOSITION C % Si % Mn % Cr % Nb % Ni, n.d.

[62] D.I. Yakubovsky, A. V. Arsenin, Y. V. Stebunov, D.Yu. Fedyanin, V.S. Volkov, Optical constants and structural properties of thin gold films, Opt. Express 25 (2017) 25574. https://doi.org/10.1364/OE.25.025574.

[63] Y. Zhang, W. Zhu, T. Borca-Tasciuc, Thermal conductivity measurements of thin films by non-contact scanning thermal microscopy under ambient conditions, Nanoscale Adv. 3 (2021) 692–702. https://doi.org/10.1039/D0NA00657B.

[64] D.R. Smith, F.R. Fickett, Low-Temperature Properties of Silver, J. Res. Natl. Inst. Stand. Technol. 100 (1995) 119. https://doi.org/10.6028/jres.100.012.

[65] Y. Bai, L. Zhao, D. Ju, Y. Jiang, L. Liu, Wide-angle, polarization-independent and dual-band infrared perfect absorber based on L-shaped metamaterial, Opt. Express 23 (2015) 8670. https://doi.org/10.1364/OE.23.008670.

[66] L. Tang, C. Dames, Effects of thermal annealing on thermal conductivity of LPCVD silicon carbide thin films, (2023).

[67] Q. Liu, G. Ding, Y. Wang, J. Yao, Thermal Performance of Micro Hotplates with Novel Shapes Based on Single-Layer SiO2 Suspended Film, Micromachines (Basel). 9 (2018) 514. https://doi.org/10.3390/mi9100514.

[68] D. Li, Y. Ruan, C. Chen, W. He, C. Chi, Q. Lin, Design and Thermal Analysis of Flexible Microheaters, Micromachines (Basel). 13 (2022) 1037. https://doi.org/10.3390/mi13071037.

[69] N. To, S. Juodkazis, Y. Nishijima, Detailed Experiment-Theory Comparison of Mid-Infrared Metasurface Perfect Absorbers, Micromachines (Basel). 11 (2020) 409. https://doi.org/10.3390/mi11040409.

[70] V. Sorathiya, Z.A. Sbeah, A. Alghamdi, A.Y. Jaffar, A.G. Alharbi, Wideband metamaterial perfect absorber using topological insulator material for infrared and visible light spectrum: a numerical approach, Sci. Rep. 15 (2025) 28514. https://doi.org/10.1038/s41598-025-14623-7.

[71] N. Liu, M. Mesch, T. Weiss, M. Hentschel, H. Giessen, Infrared Perfect Absorber and Its Application As Plasmonic Sensor, Nano Lett. 10 (2010) 2342–2348. https://doi.org/10.1021/nl9041033.

[72] O. Rasoga, D. Dragoman, A. Dinescu, C.A. Dirdal, I. Zgura, F. Nastase, A.M. Baracu, S. Iftimie, A.C. Galca, Tuning the infrared resonance of thermal emission from metasurfaces working in near-infrared, Sci. Rep. 13 (2023) 7499. https://doi.org/10.1038/s41598-023-34741-4.

[73] K. Grujić, A Review of Thermal Spectral Imaging Methods for Monitoring High-Temperature Molten Material Streams, Sensors 23 (2023) 1130. https://doi.org/10.3390/s23031130.